\documentclass[aps,prd,eqsecnum,11pt,showpacs,nofootinbib]{revtex4-1}
\usepackage{geometry}                		
\usepackage{graphicx}				
\usepackage{amssymb}
\usepackage{graphicx}
\usepackage{amssymb,amsmath}
\usepackage{amsfonts,mathrsfs}

\newcommand{\be}{\begin{equation}}
\newcommand{\ee}{\end{equation}}
\newcommand{\bea}{\begin{eqnarray}}
\newcommand{\eea}{\end{eqnarray}}
\newcommand{\bes}{\begin{subequations}}
\newcommand{\ees}{\end{subequations}}
\newcommand{\w}{\omega}

\begin{document}

\title{Low frequency gray-body factors and infrared divergences: \\
rigorous results}


\author{Paul~R.~Anderson}
\email{anderson@wfu.edu}
\affiliation{Department of Physics, Wake Forest University, Winston-Salem, North Carolina 27109, USA}
\author{Alessandro~Fabbri}
\email{afabbri@ific.uv.es}
\affiliation{Centro Studi e Ricerche E. Fermi, Piazza del Viminale 1, 00184 Roma, Italy}
\affiliation{Dipartimento di Fisica dell'Universit\`a di Bologna, Via Irnerio 46, 40126 Bologna, Italy}
\affiliation{Departamento de F\'isica Te\'orica and IFIC, Universidad de Valencia-CSIC, C. Dr. Moliner 50, 46100 Burjassot, Spain}
\affiliation{Laboratoire de Physique Th\'eorique, CNRS UMR 8627, B\^at. 210, Universit\'e Paris-Sud 11, 91405 Orsay Cedex, France}
\author{Roberto~Balbinot}
\email{Roberto.Balbinot@bo.infn.it}
\affiliation{Dipartimento di Fisica dell'Universit\`a di Bologna, Via Irnerio 46, 40126 Bologna, Italy}
\affiliation{INFN sezione di Bologna, Via Irnerio 46, 40126 Bologna, Italy}
\affiliation{Centro Studi e Ricerche E. Fermi, Piazza del Viminale 1, 00184 Roma, Italy}

\begin{abstract}
Formal solutions to the mode equations for both spherically symmetric black holes and Bose-Einstein condensate acoustic black holes are obtained by writing the spatial part of the
mode equation as a linear Volterra integral equation of the second kind.  The solutions work for a massless minimally coupled scalar field in the s-wave or zero angular momentum sector for a spherically symmetric black hole and in the longitudinal sector of a
1D Bose-Einstein condensate acoustic black hole. 
These solutions are used to obtain in a rigorous way analytic expressions for the scattering coefficients and gray-body factors in the zero frequency limit.  They are also used to study the infrared behaviors of the symmetric two-point function and two functions derived from it: the point-split stress-energy tensor for the massless minimally coupled scalar field in Schwarzschild-de Sitter spacetime and the density-density correlation function for a Bose-Einstein condensate acoustic black hole.
\end{abstract}

\maketitle

\section{Introduction}
\label{sec-intro}

A major prediction of quantum field theory in curved space is that when a black hole, BH, forms from the gravitational collapse of a star, particle production
occurs and the BH evaporates~\cite{Hawking:1974sw}.    If backreaction effects are neglected then,
after a transient which depends on the details of the collapse, the emission rapidly becomes stationary, is nearly thermal, and depends only on the geometry exterior to the star and to the event horizon of the BH.  The temperature of the radiation is proportional to the surface gravity $\kappa$ of the event horizon.
The fact that this radiation depends only on
the parameters (conserved charges) characterizing the spacetime exterior to the BH
is a manifestation of the BH no-hair theorem (see for example~\cite{hawkell}).

The spectrum of the radiated particles is not exactly thermal because the modes of the quantum fields propagating from the horizon region to infinity experience the presence of an effective potential that causes partial reflection back to the horizon. So, the number of particles which propagate to infinity\footnote{In the case of a black hole immersed in de Sitter space this
is the number of particles that propagate to the cosmological horizon.  See Sec.~\ref{sec-schds} for more details.} for each mode (in units where $k_B=\hbar=c=1$) is
\be\label{kjk}
N_\omega^{(i)}=\frac{\Gamma^{(i)}(\omega)}{e^{\frac{2\pi\omega}{\kappa}}-1} \;.  \ee
Here $\kappa$ is the surface gravity of the event horizon, $\omega$ is the frequency of the mode, and $i$ denotes the other quantum numbers needed to specify the mode.
The factor $\Gamma^{(i)}(\omega)$ is the so called `gray-body factor' which accounts for the backscattering mentioned above.
In the literature $\Gamma^{(i)}(\omega)$ is usually computed analytically by the method of asymptotic matching~\cite{teukpr, starob, page, uno, due}. Since the effective potential
vanishes at the horizon and at infinity (or the cosmological horizon) it is possible to write the solutions to the mode equation at these locations in terms of plane waves.
These are then matched in the intermediate region.
Particular interest lies in the low frequency behavior of $\Gamma^{(i)}(\omega)$ because the infrared divergence coming from the Planckian factor in~\eqref{kjk} can be avoided  if $\Gamma^{(i)}(\omega)$ vanishes in the $\omega\to 0$ limit.  If $\Gamma^{(i)}(\omega)$ approaches a nonzero constant in this limit then an infinite number of `soft' particles will be emitted.

For asymptotically flat spherically symmetric BHs one has for the $\ell=0$ mode, which is the dominant one for BH radiation, a universal behavior at low frequency $\Gamma^{(\ell=0)}(\omega)\sim A_H\omega^2$ where $A_H$ is the area of the event horizon \cite{uu, uuu}.  However for BHs in de Sitter space one has $\Gamma^{(\ell=0)}(\omega)\sim {\rm const}.$ in the low frequency limit for the massless minimally coupled scalar field~\cite{unoref, kgb}.\footnote{For nonminimal coupling to the scalar curvature or $\ell\neq 0$ one recovers even for these BHs a vanishing gray-body factor (see for example \cite{chor} and \cite{kpp}).} A nonvanishing $\Gamma^{(i)}(\omega)$ in the $\omega\to 0$ limit was also found recently~\cite{abfp2} for a class of Bose-Einstein condensate, BEC, acoustic BHs which have effectively have one spatial dimension. 
It is important to stress that the infrared divergences in the number of created particles found for the Schwarzschild de Sitter and BEC cases are
weak enough that there is no infrared divergence in the combined energies of these particles.

 A possible problem with the asymptotic matching method described above and used in~\cite{teukpr, starob, page, uno, due,abfp2} is that the near-horizon and zero frequency limits do not commute for solutions to the mode equation.  The same is true for the zero frequency and infinite distance limits (or in the Schwarzschild-de Sitter case, the zero frequency limit and the limit in which the cosmological horizon is approached).

In this paper, we provide rigorous derivations of the zero frequency limits for the scattering coefficients and gray-body factors for the mode functions of the massless minimally coupled scalar field in spacetimes containing spherically symmetric black holes and for BEC acoustic black holes.
For spherically symmetric black holes, we restrict our attention to modes with zero angular momentum, i.e. the s-wave sector.  For BEC acoustic black holes we assume that the system can be approximated as having just one space dimension and thus work in the longitudinal sector.  We also only consider cases in which the velocity $v$ of the condensate is constant and the sound speed $c$ varies as a function of position. Qualitatively similar results can be obtained for more general cases in which both $c$ and $v$ vary with position.  

   To compute the scattering coefficients and gray-body factors we have developed a general analytic method, based on a linear Volterra integral equation of the second kind~\cite{volterra-eq}, for solving the mode equation. In all cases, we compute the exact (to leading order in $\omega$) scattering coefficients in the exterior of the black hole and relate them to the horizon boundary values of the appropriate exact solutions (and their first derivatives) of the zero frequency mode equation.

    We use the results to investigate infrared divergences in the symmetric two-point function for the massless minimally coupled scalar field in the Unruh state~\cite{unruh} in both the BH and BEC cases.  To our knowledge the only prior work in this area is in~\cite{abfp2} where it was shown that an infrared divergence occurs for the two-point
    function in the BEC case.\footnote{There were also results announced in~\cite{abfp2} for the two-point functions for spherically symmetric black holes, for the point-split stress-energy tensor, and the density-density correlation function, but the proofs of these results are given here.}
    We find that the two-point function is infrared finite if the gray body factor vanishes in the zero frequency limit.  However, it is
  infrared divergent if the gray body factor does not vanish in the zero frequency limit.
  From the relationship between the two-point function and the point-split stress-energy tensor for a spherically symmetric black hole or the density-density correlation function for a BEC acoustic black hole, one would guess that infrared divergences in the two-point function would
lead to infrared divergences in these quantities as well.  We show that this is not the case and that there are no infrared divergences in either the point-split stress-energy tensor for the massless minimally coupled scalar field in Schwarzschild-de Sitter spacetime or the density-density correlation function for the class of BEC acoustic black holes that we consider.

In Sec.~\ref{sec-volterra} we find two formal linearly independent solutions to the mode equation in terms of Volterra integral equations of the second kind.  We then find a second set of formal solutions to the spatial part of the mode equation which are useful near the event horizon.  In Sec.~\ref{sec-scattering-coefficients} we use these solutions to
derive formal expressions for the scattering coefficients and gray-body factors at all frequencies.  In Sec.~\ref{sec-low-freq} it is shown that in the low frequency limit the scattering coefficients and gray-body factor can be written in terms of exact solutions to the zero frequency mode equation which are evaluated in the limit that the event horizon is approached.
The condition which determines whether the gray-body factor vanishes or approaches a nonzero constant in the zero frequency limit is given.  In Sec.~\ref{sec-examples}
specific expressions for the low frequency limits of the scattering coefficients and gray-body factor are derived for Schwarzschild, Reissner-Nordstr\"{o}m, and Schwarzschild-de Sitter spacetimes as well as for the class of BEC acoustic BHs we are considering.  In Sec.~\ref{sec-correlation} the question of infrared divergences in the two-point function, the point-split stress-energy tensor, and the density-density correlation function are investigated.    A summary and discussion of our results is given in Sec.~\ref{sec-concl}.  In Appendix A some of the details of the derivations in Sec.~\ref{sec-low-freq} are given.  In Appendix B exact (in $\omega$) near-horizon solutions to the mode equation are constructed.

\section{Volterra integral equation}
\label{sec-volterra}

The Klein-Gordon equation for a massless minimally coupled scalar field
\begin{equation}\label{kg}
\frac{1}{\sqrt{-g}}\partial_{\mu}\left(\sqrt{-g}\partial^{\mu}\,  \phi \right)=0
\end{equation}
arises both for a quantum field in curved space and as a description of the quantized phase fluctuations (phonons)  of a BEC in the hydrodynamic limit (details are given, e.g., in \cite{abfp}).
For both  an analog black hole and a gravitational black hole, the class of metrics we shall be interested in can be written in the form
\begin{equation}\label{geme}
ds^2=-f(x)dt^2+\frac{dx^2}{f(x)}+C_{\perp}^2(x) dx_{\perp}^2 \ ,
\end{equation}
where $f=0$ (and $x=0$) at the horizon, and we have explicitly separated the ($1+1$) and `transverse' parts.
For a spherically symmetric black hole
\be C_{\perp}^2(x) dx_{\perp}^2= (x+r_H)^2 d\Omega^2 \;, \ee
 with $r_H$ the radius of the event horizon.  The usual radial coordinate is $r = x + r_H$ and we use units such that $\hbar = c = G = k_B = 1$.   For a BEC it can be shown  that the original acoustic metric can be mapped, via a coordinate transformation, to the form (\ref{geme}) with
 \be C_{\perp}^2(x) dx_{\perp}^2 = \frac{n}{m c}(dy^2+dz^2) \;, \ee
 where $n$ is the number density of particles in the condensate and $m$ is the mass of an individual particle.
    Here we consider for simplicity only models in which the BEC moves with constant velocity $v<0$ in the $x$ direction, $n$ is a constant, and the sound speed $c$ varies with $x$
     (see e.\ g.\ \cite{abfp}), but our analysis can be easily extended to models in which $c,\ v$ and $n$ vary with $x$. In what follows we ignore the irrelevant constants $n$ and $m$ which is equivalent to choosing units such that $n/m = 1$.  Further in the gravitational case we consider only the spherically symmetric $l=0$ modes and in the BEC case only the longitudinal ones.  Thus in both cases $\phi=\phi(t,x)$.
For stationary solutions
\be \phi(x,t) = \phi_\w (x) e^{-i \w t}  \;, \ee
and Eq.~\eqref{kg} becomes
\be  \frac{1}{C_\perp^2(x)} \, \frac{d}{dx} \left[ C_\perp^2(x) f(x) \frac{d}{d x} \right] \phi_\w (x) + \frac{\w^2}{f(x)} \phi_\w(x) = 0  \label{mode-phi} \;. \ee

A useful alternative form for the mode equation can be derived by making the changes of variable
\bes \bea \phi(t,x) &=& \frac{\chi(t,x)}{C_\perp(x)} \;, \label{chi-def} \\
    x^* &\equiv& \int^x \frac{d\bar{x}}{f(\bar{x})} \;,  \label{xstar-def} \eea \label{chi-xstar-def} \ees
where $x^*$ is the tortoise coordinate.
To determine the behavior of $x^*$ near the horizon note that
for the cases we consider it
is possible to expand the metric coefficient $f(x)$ in~\eqref{geme} in powers of $x$ such that
\be  f(x) = \sum_{i = 1}^\infty f_i  x^i  \;.  \label{f-exp}  \ee
The definition of the surface gravity is
\be  \kappa = \frac{1}{2} \frac{d f(x)}{d x}\bigg|_{x = 0}  \; , \label{kappa-def} \ee
thus near the horizon
\be  f = 2 \kappa x + O(x^2) \;. \label{f-exp-2} \ee
Substituting into~\eqref{xstar-def} it is easy to see that near the horizon $x^* \to - \infty$ and more precisely that
\be x = x_0 e^{2 \kappa x^*} + O( e^{4 \kappa x^*} ) \;, \label{x-xstar-exp} \ee
with $x_0$ an arbitrary positive constant.
Far from the horizon, if $f\to {\rm const.}$ as $x^*\to +\infty$ then $x^*\sim x$.  If instead there is a cosmological horizon then $x^* \to + \infty$ at the cosmological horizon.

Substituting~\eqref{chi-xstar-def} into~\eqref{kg} gives
\be \left[- \partial_t^2 +  \partial_{x^*}^2  - V_{\rm eff} \right] \chi(t,x) = 0  \;. \label{cau}  \ee
Here we have introduced the effective potential
\be\label{veff}
\ \ \ V_{\rm eff}=\frac{1}{C_{\perp}} \frac{d^2 C_\perp}{d ({x^*})^2}  = \frac{1}{C_{\perp}} f \, \frac{d}{d x} \left( f \, \frac{d C_\perp}{dx} \right) \ .
\ee
For stationary solutions $\chi(x,t) = \chi_\w e^{-i \w t}$, Eq.~\eqref{cau}
 takes the Schr\"odinger-like form
\begin{equation}\label{schreq}
\left[ \frac{d^2}{d (x^*)^2} +\omega^2 - V_{\rm eff}\right] \chi_\omega =0\ .
\end{equation}

For the cases we consider $V_{\rm eff}$ vanishes at $x^* =  \pm\infty$.
Thus the general solution to Eq.~\eqref{schreq} will be a linear combination of two solutions $\chi^c_\w$ and $\chi^s_\w$ with
the asymptotic behaviors
\be   \chi^c_\omega \to \cos \omega x^* \,,  \qquad  \chi^s_\omega \to \sin \omega x^* \,, \qquad {\rm as} \qquad x^* \to \infty  \;. \label{c-s-large-x}  \ee
Formal solutions for $\chi^c_\w$ and $\chi^s_\w$ can be obtained in terms of the following linear Volterra integral equations of the second kind:
\bes \label{volterra} \bea
\chi^c_\w(x^*) &=& \cos \w x^* - \frac{1}{\omega}\int_{x^*}^{\infty}\sin[\omega(x^*-y^*)]V_{eff}(y^*)\chi^c_\w(y^*)  \;, \\
\chi^s_\w(x^*) &=& \sin \w x^* - \frac{1}{\omega}\int_{x^*}^{\infty}\sin[\omega(x^*-y^*)]V_{eff}(y^*)\chi^s_\w(y^*)  \;.
\eea  \ees

For all $x^*$ these solutions can also be written in the general form
\bea  \chi^c_\omega (x^*) &=& A(\omega, x^*) \cos \omega x^* \, +  B(\omega, x^*) \sin \omega x^*  \ , \nonumber \\
      \chi^s_\omega (x^*) &=& C(\omega, x^*) \cos \omega x^* \, +  D(\omega, x^*) \sin \omega x^*  \; ,
       \label{abcd}  \eea
 with
 \bea A(\omega, x^*) &=& 1 + \frac{1}{\omega} \int_{x^*}^\infty d y^* \, \sin \omega y^* \, V_{\rm eff}(y^*) \chi^c_\omega(y^*)  \;, \nonumber \\
     B(\omega, x^*) &=& - \frac{1}{\omega} \int_{x^*}^\infty d y^* \, \cos \omega y^* \, V_{\rm eff}(y^*) \chi^c_\omega(y^*)   \;, \nonumber \\
    C (\omega, x^*)  &=& \frac{1}{\omega} \int_{x^*}^\infty d y^* \, \sin \omega y^* \, V_{\rm eff}(y^*) \chi^s_\omega(y^*)   \;, \nonumber \\
    D (\omega, x^*) &=& 1 - \frac{1}{\omega} \int_{x^*}^\infty d y^* \, \cos \omega y^* \, V_{\rm eff}(y^*) \chi^s_\omega(y^*)  \label{abcddef} \;.\eea
We note that
\bea \frac{d\chi^c_w}{dx^*}&=&- A(\w,x^*) \w \sin \omega x^* \, + B(\w, x^*) \w \cos \omega x^*\ , \nonumber \\
\frac{d\chi^s_w}{dx^*}&=&- C(\w, x^*)\w \sin \omega x^* \, +  D(\w,x^*) \w\cos \omega x^* \;, \label{abcd-derivatives}      \eea
and that the
Wronskian
\be W = \chi_\w^* \frac{d \chi_\w}{d x^*} - \frac{d \chi_\w^*}{d x^*} \chi_\w \;, \label{wronskian} \ee
is a constant.  For $\chi_\w =  \chi^c_\omega + i \chi^s_\omega $ it can easily be seen by evaluating (\ref{wronskian}) in the large $x^*$ limit, that $ W = 2 i \omega$ and thus that
\be  \frac{1}{\w} \left[ \chi^c_\w \frac{d \chi^s_\w}{d x^*} - \chi^s_\w \frac{d \chi^c_\w}{d x^*} \right] = 1  \;. \label{wron-cos-sin} \ee
Using Eqs.~\eqref{abcd} and~\eqref{abcd-derivatives}
one finds
\be A(\omega,x^*)  D(\omega,x^*) - B(\omega,x^*)  C(\omega,x^*) = 1  \; .  \label{ABDCwronskian} \ee

\section{Scattering coefficients and grey-body factors}
\label{sec-scattering-coefficients}

 In the horizon limit, $x^* \to -\infty$, the solutions~\eqref{abcd} have the asymptotic behaviors 
 \bea  \chi^c_\omega &\to& A_\omega \cos \omega x^* \, + B_\omega \sin \omega x^*  \ , \nonumber \\
      \chi^s_\omega &\to& C_\omega \cos \omega x^* \, + D_\omega \sin \omega x^*  \; ,  \label{ABCD}  \eea
with
\bea A_\omega &\equiv& A(\omega, -\infty)= 1 + \frac{1}{\omega} \int_{-\infty}^\infty d y^* \, \sin \omega y^* \, V_{\rm eff}(y^*) \chi^c_\omega(y^*)  \;, \nonumber \\
     B_\omega &\equiv& B(\omega, -\infty)= - \frac{1}{\omega} \int_{-\infty}^\infty d y^* \, \cos \omega y^* \, V_{\rm eff}(y^*) \chi^c_\omega(y^*)   \;, \nonumber \\
    C_\omega  &\equiv& C(\omega,-\infty) = \frac{1}{\omega} \int_{-\infty}^\infty d y^* \, \sin \omega y^* \, V_{\rm eff}(y^*) \chi^s_\omega(y^*)   \;, \nonumber \\
    D_\omega &\equiv& D(\omega,-\infty)= 1 - \frac{1}{\omega} \int_{-\infty}^\infty d y^* \, \cos \omega y^* \, V_{\rm eff}(y^*) \chi^s_\omega(y^*)  \label{ABCDdef} \;.
\eea

We shall now relate the coefficients (\ref{ABCDdef}) to the scattering coefficients in the exterior of the black hole (the $f>0$ region in (\ref{geme}))  \cite{abfp}.
Because of the independence of Hawking radiation from the details of the collapse producing the BH, when discussing this emission one can consider just the spacetime
exterior to the star, mimicking the effects of the collapse by giving appropriate boundary conditions on the past horizon of the analytically extended exterior manifold.
A complete basis for the solutions of the mode equation (\ref{cau})  is formed by two sets of modes, $\chi_I$ and $\chi_H$, shown in Figs. 1, 2 \cite{candelas}.
\begin{figure}[h]
\centering \includegraphics[angle=0, height=2.0in] {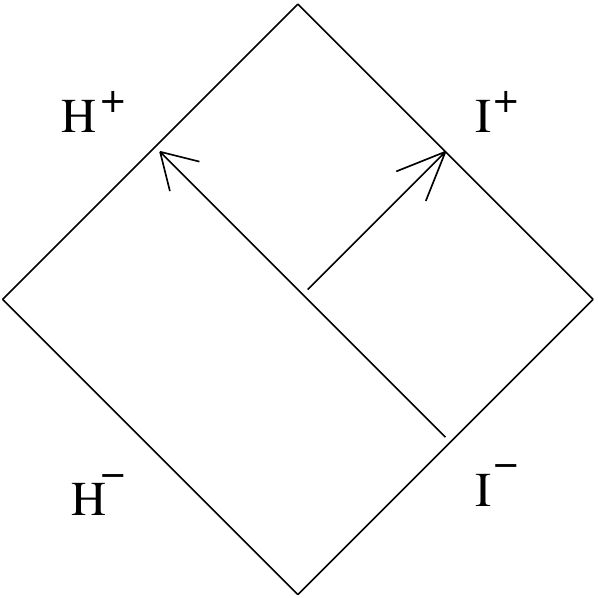}
\caption{Modes $\chi_I$, starting at past null infinity ($I^-$), transmitted to the future horizon ($H^+$) and reflected to future null infinity ($I^+$).}
\end{figure}
\begin{figure}[h]
\centering \includegraphics[angle=0, height=2.0in] {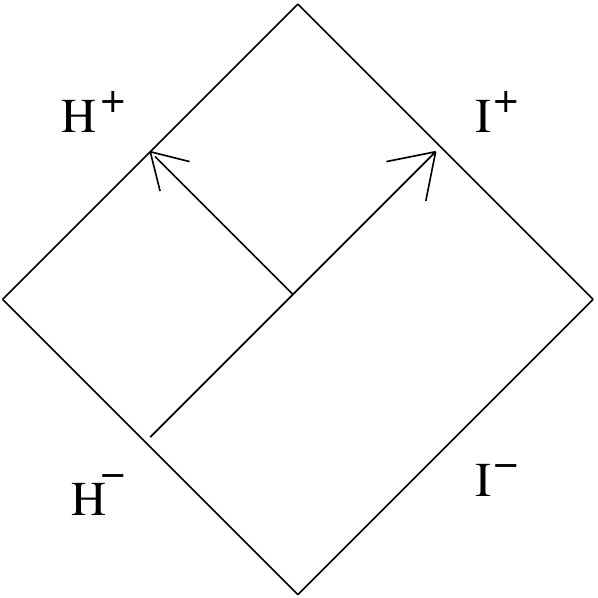}
\caption{Modes $\chi_H$, starting at the past horizon ($H^-$), transmitted to future null infinity ($I^+$) and reflected to the future horizon ($H^+$).}
\end{figure}
The modes $\chi_I$ originate at past null infinity $I^-$ ($x^*=-\infty$) and because of the potential term in Eq. (\ref{schreq}) are partially transmitted towards the future horizon $H^+$ ($x^* = +\infty$)  and partially reflected to future null infinity $I^+$ ( $x^* = +\infty$).
The modes $\chi_H$ originate on the past horizon $H^-$ ( $x^* =  -\infty$). They are partially transmitted to future null infinity $I^+$, and partially reflected back to the future horizon
$H^+$.
More specifically
\bes \bea \label{chiI}
\chi_I(t,x)  &=& \frac{N}{\sqrt{ \omega}}e^{-i\omega t}\chi_\omega^I(x)\ ,\\
\label{chiH}
\chi_H(t,x) &=& \frac{N}{\sqrt{ \omega}}e^{-i\omega t}\chi_\omega^H(x)\ , \eea \label{chiIH} \ees
with $\chi^I_\w$ and $\chi^H_\w$ solutions to Eq.~\eqref{schreq}.  For a 1D BEC acoustic black hole $N = \frac{1}{\sqrt{4 \pi}}$ and for a spherically symmetric black hole $N = \frac{1}{4 \pi}$.

It is useful to write $\chi^I_\w$ and $\chi^H_\w$ in terms of solutions to the mode equation~\eqref{schreq} which correspond to outgoing ($e^{i \w x^*}$) and incoming ($e^{-i \w x^*}$)
waves at $x^* = \infty$.  These solutions are
\be  \chi^{\infty}_r = \chi^{c}_\omega + i \chi^{s}_\omega  \;, \ \ \
           \chi^{\infty}_l = \chi^{c}_\omega- i \chi^{s}_\omega  \;. \label{chi-r-l-c-s}
\ee
They have both right and left moving parts near the horizon ($x^* \to -\infty$) so that
\bea  \chi^{\infty}_r &\rightarrow& E_r e^{i \omega x*}  + F_r e^{-i \omega x*} \;, \nonumber \\
         \chi^{\infty}_l &\rightarrow& E_l e^{i \omega x*}  + F_l e^{-i \omega x*} \;, \label{chi-RL-x-0} \eea
where, using (\ref{ABCD}),
\bea  E_r &=&  \frac{1}{2} [ A_\omega+ D_\omega - i(B_\omega - C_\omega)] \;, \nonumber  \\
           F_r &=&  \frac{1}{2} [ A_\w - D_\w + i(B_\w + C_\w)] \;,  \nonumber \\
           E_l &=&  \frac{1}{2} [ A_\w - D_\w - i(B_\w + C_\w)] \;,  \nonumber \\
           F_l &=&  \frac{1}{2} [ A_\w + D_\w + i( B_\w - C_\w )] \ .
 \label{EFABCD} \eea

 To construct $\chi^I_\w$ we consider a mode corresponding to an initial unit amplitude left-moving wave at $x^*\to +\infty$ which gets partially reflected back to $x^*\to +\infty$ and partially transmitted to the horizon ($x^*\to -\infty$).  Then
\be \chi_\omega^I = \chi^{\infty}_l+R_I\chi^{\infty}_r   \rightarrow  e^{-i\omega x^*}+R_I e^{i\omega x^*}  \;, \;\;\; {\rm as} \;\;\; x^* \rightarrow +\infty  \;, \label{chi-I-inf} \;. \ee
Near the horizon this solution has the behavior
\be \chi_\omega^I \rightarrow  T_I e^{-iwx^*}   \;. \label{chi-I-H} \ee
We find, from (\ref{chi-RL-x-0}) and (\ref{EFABCD})  that
\bea
R_I=-\frac{E_l}{E_r}=-\frac{( A_\w- D_\w - i( B_\w + C_\w ))}{( A_\w +D_\w - i( B_\w - C_\w))}\ ,\nonumber \\
T_I=F_l - \frac{E_lF_r}{E_r}=\frac{1}{E_r}=\frac{2}{(A_\w+ D_\w-i( B_\w- C_\w))}\ , \label{RTB}\eea
where in the last equation we have used the Wronskian condition (\ref{ABDCwronskian}).

To construct $\chi^H_\w$ we start with a mode corresponding to a unit amplitude right-moving wave which originates at the horizon ($x^*\to-\infty$) and is partially reflected back to $x^*\to -\infty$ and partially transmitted to $x^*\to +\infty$.
   Thus
   \be \chi_\omega^H =  T_H \chi^{\infty}_r \rightarrow T_H e^{iwx^*} \;\;\; {\rm as} \;\;\; x^* \rightarrow +\infty \;.  \label{chi-H-inf} \ee
Near the horizon this solution has the behavior
   \be \chi_\omega^H  \rightarrow e^{i\omega x^*}+R_H e^{-i\omega x^*} \;. \label{chi-H-H} \ee
Thus we find, from (\ref{chi-RL-x-0}) and (\ref{EFABCD})  that
\bea R_H=\frac{F_r}{E_r}=\frac{( A_\w- D_\w + i( B_\w + C_\w))}{( A_\w+ D_\w - i( B_\w - C_\w ))}\ ,\nonumber \\
T_H=\frac{1}{E_r}=\frac{2}{(A_\w+ D_\w-i( B_\w- C_\w))}\ . \label{RTH}\eea

Finally, the gray-body factor is defined as
\be\label{grb} \Gamma \equiv |T_H|^2 \ee
and represents the probability that a mode which originates on the past horizon $H^-$ and has unit norm reaches $I^+$ ( $x^* = +\infty$).
It appears in Eq.~(\ref{kjk}) where it modifies the exact black-body spectrum for the number of particles which reach $I^+$ at a given frequency $\w$.

\section{Low-frequency limit}
\label{sec-low-freq}

In this section we consider the low-frequency limits of the expressions derived in the previous subsections.
Therefore we need to analyze, the behaviors of the integrals (\ref{ABCDdef}) for small values of $\omega$.
 The main difficulty with evaluating these integrals in the limit $\omega \rightarrow 0$ is that this limit does not commute with either of the limits $x^* \rightarrow \pm \infty $ and it is not possible to compute the integrals analytically for arbitrary values of $\omega$.  However, one can break the integrals into three parts so that
 \be  \int_{-\infty}^\infty d y^*  \;\; \rightarrow \int_{-\infty}^{- \Lambda}  d y^*  \; \; +  \int_{-\Lambda}^\Lambda  dy^* \; \;  + \int_\Lambda^\infty dy^*  \;\;, \label{3-integrals} \ee
 and take the limits $\omega \rightarrow 0$ and $\Lambda \rightarrow \infty$ in such a way that $\omega \Lambda \ll 1$.  Then expansions can be used near the horizon and at large $y^*$ for the potential $V_{\rm eff}$ in the first and third integrals which make it possible to compute these integrals analytically.  It is shown in Appendix A that when this is done these
 integrals vanish for all values of $\omega$ including $\omega = 0$ provided $V_{\rm eff}$ approaches zero fast enough at the horizon and at infinity.  It is also shown in Appendix A
 that for the middle integral we can use Eq.~\eqref{schreq} to obtain the relation $V_{\rm eff}(y^*)\chi_\omega (y^*)=[\partial_{y^*}^2 +\omega ^2]\chi_w(y^*)$, integrate twice by parts, and
 obtain
\bea A_\omega &=& 1 + \lim_{\Lambda\to \infty} \left[ \left(\frac{\sin(\omega y^*)}{\omega}\frac{d}{d y^*} -\cos(\omega y^*)\right)\chi_\omega^c(y^*) \right]_{-\Lambda}^{\Lambda}   \;, \nonumber \\
     B_\omega &=& -  \lim_{\Lambda\to \infty}\left[ \left(\frac{\cos(\omega y^*)}{\omega}\frac{d}{d y^*} +\sin(\omega y^*)\right)\chi_\omega^c (y^*)\right]_{-\Lambda}^{\Lambda}    \;, \nonumber \\
    C_\omega  &=&  \lim_{\Lambda\to \infty}\left[ \left(\frac{\sin(\omega y^*)}{\omega}\frac{d}{d y^*} -\cos(\omega y^*)\right)\chi_\omega^s (y^*) \right]_{-\Lambda}^{\Lambda}    \;, \nonumber \\
   D_\omega &=& 1 - \lim_{\Lambda\to \infty}\left[ \left(\frac{\cos(\omega y^*)}{\omega}\frac{d}{d y^*} +\sin(\omega y^*)\right)\chi_\omega^s (y^*)\right]_{-\Lambda}^{\Lambda}  \label{ABCDdefstep1} \;.
\eea

Using the asymptotic behaviors (\ref{c-s-large-x})  it is easy to show
that in the limit $\Lambda \rightarrow \infty$ these coefficients only depend on the parts of the above expressions which are evaluated at $y^* = -\Lambda$.  Thus
\bea A_\omega &=&   \lim_{\Lambda \rightarrow \infty} \left[   \left(\cos(\omega y^*)- \frac{\sin(\omega y^*)}{\omega}\frac{d}{d y^*} \right)\chi_\omega^c (y^*) \right]_{y^*=-\Lambda}   \;, \nonumber \\
     B_\omega &=&   \lim_{\Lambda \rightarrow \infty} \left[ \left(\frac{\cos(\omega y^*)}{\omega}\frac{d}{d y^*} +\sin(\omega y^*)\right)\chi_\omega^c(y^*) \right]_{y^*=-\Lambda}    \;, \nonumber \\
    C_\omega  &=&  \lim_{\Lambda \rightarrow \infty} \left[ \left(\cos(\omega y^*)-\frac{\sin(\omega y^*)}{\omega}\frac{d}{d y^*} \right)\chi_\omega^s(y^*) \right]_{y^*=-\Lambda}   \;, \nonumber \\
  D_\omega &=& \lim_{\Lambda \rightarrow \infty} \left[ \left(\frac{\cos(\omega y^*)}{\omega}\frac{d}{d y^*} +\sin(\omega y^*)\right)\chi_\omega^s(y^*) \right]_{y^*=-\Lambda}    \label{ABCDdeffin} \;.
\eea
Recalling that we take the $\w \to 0$ and $\Lambda \to \infty$ limits in
such a way that $\w \Lambda \ll 1$, we define the following constants:
  \bea \mathscr{A} &\equiv& A_0 = \lim_{\stackrel {\omega \to 0}{\Lambda\to\infty}}   \left[ \left(\cos(\omega y^*)- \frac{\sin(\omega y^*)}{\omega}\frac{d}{d y^*} \right)\chi_\omega^c(y^*) \right]_{y^*=-\Lambda}   \;, \nonumber \\
   \mathscr{B} &\equiv&  \lim_{\omega \to 0} (\omega B_\w ) =  \lim_{\stackrel {\omega \to 0}{\Lambda\to\infty}}  \left[ \left(\frac{\cos(\omega y^*)}{\omega}\frac{d}{d y^*} +\sin(\omega y^*)\right) \omega \chi_\omega^c(y^*) \right]_{y^*=-\Lambda}    \;, \nonumber \\
   \mathscr{C} &\equiv& \lim_{\omega \to 0} (\w^{-1} C_\w )  =  \lim_{\stackrel {\omega \to 0}{\Lambda\to\infty}} \left[ \left(\cos(\omega y^*)-\frac{\sin(\omega y^*)}{\omega}\frac{d}{d y^*} \right) \frac{\chi_\omega^s(y^*)}{\omega} \right]_{y^*=-\Lambda}    \;, \nonumber \\
  \mathscr{D} &\equiv&  D_0 =  \lim_{\stackrel {\omega \to 0}{\Lambda\to\infty}} \left[ \left(\frac{\cos(\omega y^*)}{\omega }\frac{d}{d y^*} +\sin(\omega y^*)\right)\chi_\omega^s(y^*)\right]_{y^*=-\Lambda}    \label{ABCDsmallw} \;.
\eea
 Using the Wronskian condition~\eqref{wron-cos-sin} it is easy to show that
 \be \mathscr{A} \mathscr{D}-\mathscr{B} \mathscr{C} =1 \;. \label{wron-scriptABCD} \ee
We find\footnote{The relation for $A_\w$ is proved in Appendix A.  The proofs of the other relations are straight-forward generalizations of that proof.} that for small $\w$
\bea A_\w &=& \mathscr{A} + O(\w^2) \;, \nonumber \\
     B_\w &=& \frac{\mathscr{B}}{\w} + O(\w)  \;, \nonumber \\
     C_\w &=& \w \mathscr{C}  + O(\w^3)  \;, \nonumber \\
     D_\w &=& \mathscr{D} + O(\w^2)  \;. \label{ABCD-scrip-ABCD} \eea

To evaluate the expressions in~\eqref{ABCDsmallw} we first note that since $\w \Lambda \ll 1$
\bea \mathscr{A} &=& \lim_{\stackrel {\omega \to 0}{\Lambda\to\infty}}   \left[ \left(1-  y^* \frac{d}{d y^*} \right)\chi_\omega^c(y^*) \right]_{y^*=-\Lambda}   \;, \nonumber \\
   \mathscr{B} &=&   \lim_{\stackrel {\omega \to 0}{\Lambda\to\infty}}  \left[ \left(\frac{d}{d y^*}\right) \chi_\omega^c(y^*) \right]_{y^*=-\Lambda}    \;, \nonumber \\
   \mathscr{C} &=&  \lim_{\stackrel {\omega \to 0}{\Lambda\to\infty}} \left[ \left(1-  y^* \frac{d}{d y^*} \right) \frac{\chi_\omega^s(y^*)}{\omega} \right]_{y^*=-\Lambda}    \;, \nonumber \\
  \mathscr{D} &=&  \lim_{\stackrel {\omega \to 0}{\Lambda\to\infty}} \left[ \frac{d}{d y^*} \frac{\chi_\omega^s(y^*)}{\omega} \right]_{y^*=-\Lambda}    \label{ABCDsmallw2} \;.
\eea
Next we show that both $\chi_\omega^c(x^*)$ and $\frac{\chi_\omega^s(x^*)}{\omega}$ and their first derivatives can be evaluated in the limit $\w \to 0$ with  $\w |x^* |\ll 1$
by identifying them with solutions to the mode equation when $\w = 0$.  
  To do so we consider small values of $\w$ and large values of $x^*$ with two constraints.  One is that $\w^2 \gg V_{\rm eff}(x^*)$ which means that to leading order
   $\chi^c_\w = \cos(\w x^*)$ and $\chi^s_\w = \sin(\w x^*) $.  The second is that $0 < \w x^* \ll 1$ which means that to leading order $\chi^c_\w = 1$ and $\chi^s_\w = \w x^*$.
   Examination of the large $x^*$ behaviors given in Appendix A for $V_{\rm eff}$ in the cases we
    are interested in shows that it is not hard to find values of $\w$ and $x^*$ that satisfy these two conditions.
    We want to match $\chi^c_\w$ and $\chi^s_\w$ to solutions to the mode equation~\eqref{schreq} with $\w = 0$.
  We will denote the solution that $\chi^c_\w$ matches to as $\chi^{(1)}_0 $.
    The situation for $\chi^s_\w$ is more subtle because $\chi^s_\w \approx \w x^*$  and thus vanishes in the limit
   $\w \to 0$. However, $\frac{\chi^s_\w}{\w} \approx x^*$ does not vanish in this limit.  Further this quantity obeys the same mode equation as $\chi^s_\w$ and in the small $\w$ limit this is just the zero frequency mode equation.  Therefore we can match the large $x^*$ behavior of $\frac{\chi^s_\w}{\w}$ with the corresponding behavior of a solution to the zero frequency mode equation which we will denote by $\chi^{(2)}_0$.
   Thus, in the limit $x^* \to \infty$ these solutions have the behaviors
   \bes
   \bea  \chi^{(1)}_0(x^*) &\to& 1 \;, \label{chi0-1}  \\
         \chi^{(2)}_0(x^*) &\to& x^* \;.  \label{chi0-2} \eea \label{chi0-1-2} \ees
   To finish the matching we need a condition on the derivatives of these solutions.
   This can be obtained from the Wronskian condition~\eqref{wron-cos-sin} satisfied by $\chi^c_\w$ and $\chi^s_\w$.  The corresponding condition for $\chi^{(1,2)}_0$ is
 \be \chi_0^{(1)} \frac{d \chi_0^{(2)}}{d x^*} - \chi_0^{(2)} \frac{d \chi_0^{(1)}}{d x^*} = 1 \;.  \label{wron-1-2} \ee
 This fixes the asymptotic behaviors of their derivatives.
 Then we can write for $\w \to 0$ with $\w |x^*| \ll 1$
 \bea  \lim_{\w \to 0}  \chi^c_\w &=& \chi^{(1)}_0  \nonumber \\
       \lim_{\w \to 0}  \frac{\chi^s_\w}{\w} &=& \chi^{(2)}_0 \;. \label{chics-chi-1-2} \eea
 For future reference we note that therefore
 \bea  \lim_{\w \to 0}  \chi^\infty_r &=& \chi^{(1)}_0 + i \w \chi^{(2)}_0 \nonumber \\
       \lim_{\w \to 0}  \chi^\infty_\ell &=& \chi^{(1)} - i \w \chi^{(2)}_0 \;. \label{chirell-chi-1-2} \eea

 Replacing $\chi^c_\w$ and $\chi^s_\w/\w$ with the corresponding solutions $\chi^{(1,2)}_0$ in Eq.~\eqref{ABCDsmallw2} gives
  \bea \mathscr{A} &=& \left[ \left(1-  x^* \frac{d}{d x^*} \right)\chi^{(1)}_0(x^*) \right]_{hor}   \;, \nonumber \\
   \mathscr{B} &=&   \left[ \left(\frac{d}{d x^*}\right) \chi^{(1)}_0(x^*) \right]_{hor}    \;, \nonumber \\
   \mathscr{C} &=&   \left[ \left(1-  x^* \frac{d}{d x^*} \right) \chi^{(2)}_0 (x^*) \right]_{hor}    \;, \nonumber \\
  \mathscr{D} &=&    \left[ \frac{d}{d x^*} \chi^{(2)}_0(x^*) \right]_{hor}    \label{ABCDsmallw3} \;,
\eea
where the subscript ``hor'' means that the quantities inside the square brackets are to be evaluated on the horizon.  Note that the Wronskian condition~\eqref{wron-1-2} implies that
\be
\mathscr{A} \mathscr{D} - \mathscr{B} \mathscr{C} = 1  \;.  \label{wron-script-ABCD} \ee

Substituting~\eqref{ABCDdeffin} into~\eqref{RTB} and~\eqref{RTH} and using~\eqref{ABCD-scrip-ABCD} one can see that at low frequency the coefficient $\mathscr{B}$ plays a crucial role.
If $\mathscr{B}\neq 0$ then
\be\label{bd} R_I, R_H=-1+O(\omega),\ \ T_I= T_H =\frac{2i\omega}{\mathscr{B}}+O(\omega^2),\ee
while if $\mathscr{B}=0$ we have the very different behaviors
\be\label{bz} R_I=\frac{\mathscr{D}-\mathscr{A}}{\mathscr{A}+ \mathscr{D}}+O(\omega),\ \ R_H=\frac{\mathscr{A}- \mathscr{D}}{\mathscr{A}+\mathscr{D}}+O(\omega),\ \ T_I=T_H=\frac{2}{\mathscr{A}+ \mathscr{D}}+O(\omega) \ .\ee
Similarly the gray-body factor (\ref{grb}) vanishes at low-frequency as
 \be\label{grbd} \Gamma \to \frac{4\omega^2}{\mathscr{B}^2}\ee
 when $\mathscr{B}\neq 0$, while it approaches the constant value
 \be\label{grbz} \Gamma=\frac{4}{(\mathscr{A}+ \mathscr{D})^2}\ee
 when $\mathscr{B}=0$.

   From Eq.~\eqref{ABCDsmallw3} it is clear that $\mathscr{B} = 0$ if $d \chi^{(1)}_0/d x^*$ vanishes at the horizon.  The  mode
   equation~\eqref{schreq} can easily be solved exactly when $\w = 0$ by going back to Eq.~\eqref{mode-phi}, setting $\w = 0$ and
   using the relation $\phi_0 = \chi_0/C_\perp $.  The general solution is
   \be \chi_0(x)  = a \, C_\perp(x) + b \, C_\perp(x) \int^x \frac{d \bar{x}}{C_\perp^2(\bar{x}) \, f(\bar{x}) }  \;, \label{chi0-general} \ee
   with $a$ and $b$ arbitrary constants.  Then
   \be \frac{d \chi_0}{d x^*} = \frac{1}{C_\perp} \, \frac{d C_\perp}{d x}  \, f\chi_0 + \frac{b}{C_\perp}  \;. \label{dchi0dxstar} \ee
  Since for the models we consider, $f$ vanishes at the horizon and $C_\perp$ and its first derivative are nonzero constants there, it is clear that $d \chi^{(1)}_0/dx^*$ can
  only vanish at the horizon if $b = 0$.  However, $\chi^{(1)}_0$ is specified by the condition~\eqref{chi0-1} which can only be satisfied with $b = 0$ if $C_\perp$ approaches
  a constant in the limit $x^* \to \infty$.  As shown below this occurs for Schwarzschild-de Sitter spacetime and for a 1D BEC acoustic black hole.  It does not occur
  for any spherically symmetric black hole spacetime that is asymptotically flat.

  In cases where the solution $\chi^{(1)}_0$ has $b = 0$, it is necessary that $V_{\rm eff}$ changes sign at some finite value of $x^*$.  To see
  this note that for the cases we consider near the horizon
  \be C_\perp (x) = C_\perp(0) + \frac{d C_\perp(x)}{d x} \bigg|_{x=0} \, x + O(x^2) \;. \label{Cperp-exp} \ee
  Using~\eqref{x-xstar-exp} it is easy to see that near the horizon $d C_\perp/d x^*$ and $d^2 C_\perp/d (x^*)^2$ have the same sign.  Thus
  for $C_\perp$ to approach a constant value as $x^* \to \infty$ it is necessary that $d^2 C_\perp/d (x^*)^2$ changes sign.  Then Eq.~\eqref{veff} shows
  that $V_{\rm eff}$ must also change sign.

\section{Specific Cases}
\label{sec-examples}

In the  following subsections we provide specific examples of the two different behaviors which can occur for the scattering coefficients and
the gray-body factor in the low frequency limit.

\subsection{Schwarzschild and Reissner-Nordstr\"om spacetimes}
\label{sec-sch-rn}

The first, obvious, application is Schwarzschild spacetime, which will allow us to compare the results of our general procedure with those in the literature \cite{page}.
Referring to the generic metric (\ref{geme}), we have with $r = x+ r_H$, $f=1-\frac{r_H}{r}= \frac{x}{x+r_H}$ and $C_{\perp}=r = x +r_H$.  Here $r_H = 2 M$ is the horizon radius and $M$ is the mass of the black hole.  The tortoise coordinate~\eqref{xstar-def} is  $x^* =\int \frac{dx}{f} = x + r_H + r_H \ln (x/r_H) $.

  For this case, Eq. (\ref{schreq}) is the mode equation for modes in the s-wave sector of a massless scalar field.\footnote{For both Schwarzschild and Reissner-Nordstr\"om spacetimes
    the scalar curvature $R$ is zero so the mode equation is independent of the coupling to the scalar curvature.}  As illustrated in Fig. 3, the effective potential (\ref{veff})
    \be V_{\rm eff}=(1-\frac{r_H}{r})\frac{r_H}{r^3} = \frac{x \, r_H}{(x+r_H)^4} \;, \ee
     has the shape of a positive barrier, i.e. it starts from $0$ at $x=0$ ($x^*=-\infty$), reaches a maximum for $x=\frac{1}{3}r_H$ and then goes to $0$ asymptotically
 as $x, x^* \to \infty$.  The fact that it does not change sign is related to the fact that $C_\perp$ is unbounded in the limit $x^* \to \infty$.  From the discussion
 near the end of Sec.~\ref{sec-low-freq} this means that $\mathscr{B} \ne 0$ and thus the gray-body factor vanishes in the limit $\w \to 0$.

\begin{figure}
\vskip -0.2in \hskip -0.4in
\begin{center}
\includegraphics[scale=0.3,angle=90,width=3.4in,clip]{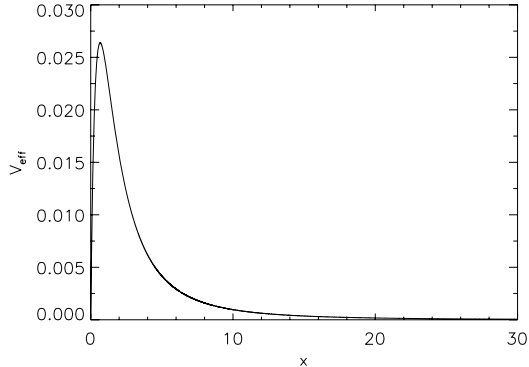}
\end{center}
\vskip -.2in \caption{Effective potential in Schwarzschild spacetime ($M=1$) plotted as a function of $x=r-r_H$.  }
\label{fig-Veff-Sch}
\end{figure}

To calculate the scattering coefficients and gray-body factor we need to find the solutions $\chi^{(1,2)}_0$ to the zero frequency mode equation.
The general solution to this equation~\eqref{chi0-general} in this case is
\begin{equation}
\chi_{0}=a (x+r_H) +b\frac{(x+r_H)}{r_H}\ln \left(\frac{x}{x+r_H} \right) .
\end{equation}
The boundary condition~\eqref{chi0-1} for $\chi^{(1)}_0$ is satisfied if
$a=0$ and $b=-1$ so that
\begin{equation}\label{phic}
\chi_{0}^{(1)}=-\frac{(x+r_H)}{r_H}\ln \left(\frac{x}{x+r_H} \right)\ .
\end{equation}
 Thus $\chi_{0}^{(1)}$ diverges as $-\frac{x^*}{r_H}$ at the horizon.

  For $\chi_{0}^{(2)}$  the boundary condition~\eqref{chi0-2} fixes $a=1$ but leaves $b$ undetermined so that
\begin{equation}\label{phis}
\chi_0^{(2)}= x + r_H +b\frac{(x+r_H)}{r_H}\ln \left(\frac{x}{x+r_H} \right) \ .
\end{equation}
Note that the Wronskian condition~\eqref{wron-1-2} is satisfied for arbitrary values of $b$.
Thus $\chi_{0}^{(2)} \sim r_H+\frac{b}{r_H}x^*$ near the horizon.
We show below that the ambiguity in the value of $b$ does not affect the low-frequency behaviors of the scattering coefficients.

 Using these explicit solutions it is easy to compute the coefficients $\mathscr{A},\ \mathscr{B},\ ,\mathscr{C},\ \mathscr{D}$  in (\ref{ABCDsmallw3}).  The result is
\bes \begin{eqnarray}\label{uno}
&& \mathscr{A}=   1 \ ,\\ \label{due}
 && \mathscr{B} = -\frac{1}{r_H}\ ,\\ \label{tre}
&&\mathscr{C} = r_H-b\ ,\\  \label{quattro}
&& \mathscr{D} = \frac{b}{r_0} \ .
\end{eqnarray} \ees

Since $\mathscr{B} \ne 0$, one finds from~\eqref{bd} that
\be\label{z}
R_I, R_H \sim -1 + O(\omega)\ , \ T_I=T_H\sim -2ir_H\omega + O(\omega^2) \;. \ee
To leading order the gray-body factor is
\begin{equation}\label{trgrsc}
\Gamma\sim 4r_H^2 \omega^2\ ,
\end{equation}
in agreement with the results in \cite{page}. We see immediately the important role played by the gray-body factor, which prevents waves with frequency $\omega \lesssim \frac{1}{2r_H}$ from reaching infinity.

 This analysis can be easily extended to Reissner-Nordstr\"om spacetimes, where the results are qualitatively similar to those for Schwarzschild spacetime.
In this case
\be f=\left(1-\frac{r_+}{x+r_+} \right)\, \left(1-\frac{r_-}{x+r_+} \right) \;, \ee
  with with $Q$ the electric charge of the black hole, $r_H = r_+=M+\sqrt{M^2-Q^2}$  the outer (event) horizon, and $r_-=M-\sqrt{M^2-Q^2}$  the inner (Cauchy) horizon.  The tortoise coordinate~\eqref{xstar-def} is  
  \be  x^* = x + r_+  - (r_+ + r_-) \ln(r_+ - r_-) + \frac{r_+^2}{(r_+-r_-)}\ln(x)-\frac{r_-^2}{r_+-r_-}\ln(x+r_+ - r_-) \;. \ee
 The general solution~\eqref{chi0-general} of the zero frequency mode equation is
\be
\chi_0=a (x+r_+) +b \left(\frac{x+r_+}{r_+-r_-} \right) \ln \left(\frac{x}{x+r_+ - r_-} \right) \;. \ee
Imposing the conditions~\eqref{chi0-1-2} one finds that
\bea
\chi^{(1)}_0 &=&  -\frac{x+r_+}{r_+-r_-}\ln \left(\frac{x}{x+r_+ -r_-}\right)  \ , \nonumber  \\
 \chi_0^{(2)}&=& x+r_+ + b\frac{x+r_+}{r_+-r_-}\ln \left(\frac{x}{r-r_-} \right) \ .
\eea
Substituting into~\eqref{ABCDsmallw3} gives the same expressions for $\mathscr{A}$, $\mathscr{B}$, $\mathscr{C}$, and $\mathscr{D}$ as in the Schwarzschild case
except that the event horizon is at $r_H = r_+$ and $r_+ \ne 2M$ unless $Q = 0$.  Similarly the expressions for the scattering coefficients and the gray-body factor are the same as for Schwarzschild to leading order.

\subsection{Schwarzschild-de Sitter spacetime}
\label{sec-schds}

We next consider a case which differs qualitatively from the previous ones, namely a black hole immersed in an expanding de Sitter universe and described by the Schwarzschild-de Sitter solution, with the metric coefficients in~(\ref{geme}) given by $f=1-\frac{2M}{r}-\frac{\Lambda}{3}r^2$ and $C_{\perp} = r $. Here $\Lambda$ is the cosmological constant.  For $0 < \Lambda < 9M^2$ the space-time has two horizons at $r = r_H$ and $r = r_C$ (with $r_H < r_C$) which are, respectively, the black hole and cosmological horizons. The equation $f=0$ has an additional negative solution ($r=-r_0<0$) which is unphysical.  We define the coordinate $x$ as before so that $x = r - r_H$.

\begin{figure}
\vskip -0.2in \hskip -0.4in
\begin{center}
\includegraphics[scale=0.3,angle=90,width=3.4in,clip]{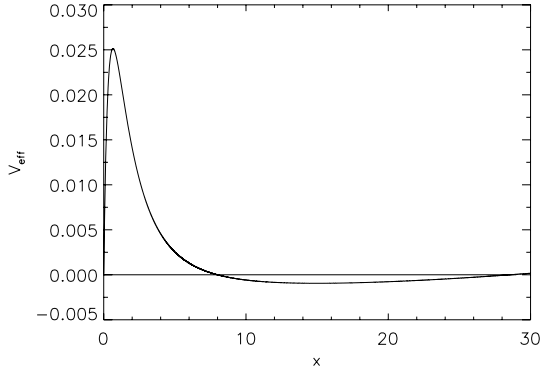}
\end{center}
\vskip -.2in \caption{Effective potential in Schwarzschild-de Sitter spacetime ($M=1$ and $\Lambda=3\times 10^{-3}$) plotted as a function of $x=r-r_H$.  }
\label{fig-Veff-SchdS}
\end{figure}

 We shall consider the mode equation (\ref{schreq}) in the region between the two horizons, $0\le x \le r_C-r_H$, which is the region where $t$ is timelike and $x$ is spacelike. 
  From Eq.~\eqref{xstar-def} it can be seen that the tortoise coordinate has the values $x^* = -\infty$ at the event horizon ( $x = 0$)  and $x^* = +\infty$ at the cosmological
   horizon ($x = r_C-r_H$).  $V_{\rm eff}$ is plotted in Fig.~\ref{fig-Veff-SchdS}.  In addition to the barrier which was also present in the Schwarzschild case, there is
     a well outside the barrier in which $V_{\rm eff} < 0$.  Since $C_\perp = r_C$ at $x^* = \infty$, the gray-body factor does not vanish in the zero frequency limit.

To compute the low-frequency scattering coefficients and gray-body factor in this spacetime we use~\eqref{chi0-general} and~\eqref{chi0-1-2} 
to find the mode functions $\chi^{(1,2)}_0$. 
Since the term proportional to $b$ in (\ref{chi0-general}) diverges as $\ln|(x + r_H -r_C)/r_C|$, the boundary condition~\eqref{chi0-1} is satisfied if   $a=\frac{1}{r_C}$ and $b=0$, so that
\be\label{ua}
\chi^{(1)}_0=\frac{x+r_H}{r_C}\ .
\ee
At the cosmological horizon the surface gravity is $\kappa_C= \frac{1}{2} |f'(x)|_{x = r_C - r_H}$ and
\be x^*\to \frac{1}{2\kappa_C} \ln \left[ \frac{|x-(r_C-r_H)|}{r_C} \right]  \;, \ee
Near the cosmological horizon, the term proportional to $b$  in (\ref{chi0-general}) has the leading order behavior 
\be  \frac{b}{2r_C\kappa_C}\ln \left[\frac{|x-(r_C-r_H)|}{r_C} \right]  \;.  \ee
 Thus the condition~\eqref{chi0-2} is satisfied if $b=r_C$ and
\be \label{ub} \chi^{(2)}_0 = a (x+r_H) +r_C (x+r_H) \int\frac{dx}{(x+r_H)^2 f}\ .\ee
Substituting these solutions into Eq.~(\ref{ABCDsmallw3}) we obtain
\bes
\begin{eqnarray}\label{l}
&& \mathscr{A}=  \frac{r_H}{r_C} \ ,\\ \label{ll}
 && \mathscr{B} = 0 \ ,\\ \label{lll}
&& \mathscr{C}= ar_H\ ,\\  \label{llll}
&& \mathscr{D} = \frac{r_C}{r_H} \ .
\end{eqnarray}
\label{ABCD-SdS}
\ees
 Since $\mathscr{B}=0$ and $\mathscr{A} \mathscr{D} =1$, the Wronskian condition~\eqref{wron-script-ABCD} is satisfied. Substituting~\eqref{ABCD-SdS} into~\eqref{bz} and~\eqref{grbz} gives
\be\label{ga}  R_I=\frac{r_C^2-r_H^2}{r_C^2+r_H^2}+O(\omega )\ , \ R_H=\frac{r_H^2-r_C^2}{r_C^2+r_H^2}+O(\omega )\ , T_I=T_H=\frac{2r_Cr_H}{r_C^2+r_H^2}+O(\omega)\
\ee
and
\be\label{vbn}
\Gamma = \frac{4r_C^2r_H^2}{(r_C^2+r_H^2)^2}\ ,
\ee
in agreement with the results of \cite{kgb}.

 In Sec.~\ref{sec-low-freq} it was shown that a necessary condition for the gray-body factor to approach a nonzero constant in the zero frequency limit is that
 $V_{\rm eff}$ should change sign somewhere between $-\infty < x^* < \infty$.  However, this is not a sufficient condition as we show next by considering
a massless scalar field with coupling $\xi \ne 0$ to the scalar curvature.  It was shown in \cite{chor} that for the $l=0$ mode
       in Schwarzschild-de Sitter spacetime $\Gamma \sim \omega^2$.  Using the fact that in Schwarzschild-de Sitter spacetime $R = 4 \Lambda$, the Klein-Gordon equation takes the form
  \begin{equation}\label{kg-xi}
\frac{1}{\sqrt{-g}}\partial_{\mu}(\sqrt{-g}\partial^{\mu}(\phi)) - \xi 4 \Lambda  \phi   =0  \;.
\end{equation}
  Transforming to $\chi = \phi/(x+r_H)$ one finds that the effective potential is
        $V_{\rm eff}=f(\frac{f'}{(x+r_H)}+4\xi\Lambda)$.  For small enough values of $|\xi|$ its qualitative behavior is the same as in Fig. 4 (see Fig. 1 of \cite{chor}).

Because $V_{\rm eff}$ approaches zero when $\xi \ne 0$ at the horizons in essentially the same way as it does when $\xi = 0$,  
 the results in Sec.~\ref{sec-low-freq} and Appendix A hold. 
For $\xi \ne 0$, the zero frequency mode equation (see~\eqref{mode-phi}) is
 \begin{equation}\label{gg-xi}
\frac{ d}{d x}\, \left[ (x+r_H)^2 f \frac{d}{d x} \left(\frac{\chi_{0}}{x+r_H} \right) \right] - 4\xi \Lambda (x+r_H) \chi_0 =0.
\end{equation}
    The crucial feature one loses when considering $\xi \neq 0$ is the existence of a solution for $\chi_0^{(1)}$ which is bounded at $x^* = \pm \infty$
  and which, in terms of the original field variable $\phi = \chi/(x+r_H)$,  corresponds to a constant field configuration \cite{unoref, dueref}.
  This can be seen for small values of $\xi$ by substituting $\chi_0 = \chi^{(1)}_0 = \chi_{0,0} + \xi \chi_{0,1}$ into~\eqref{gg-xi}.  Then to zeroth order in $\xi$ one finds 
  the general solution~\eqref{chi0-general}.  If $\chi^{(1)}_0$ is to be regular at both horizons then $\chi_{0,0}$ and $\chi_{0,1}$ must both be regular at both horizons. 
  Examination of~\eqref{chi0-general} shows that this only occurs for $\chi_{0,0}$ if $b = 0$ so that $\chi_{0,0} = a (x+r_H)$.  To first order in $\xi$ the equation is then
  \be
  \frac{ d}{d x}\, \left[ (x+r_H)^2 f \frac{d}{d x} \left(\frac{\chi_{0,1}}{x+r_H} \right) \right] = 4\xi \Lambda a (x+ r_H)^2    \;.  \label{gg-chi-1}
  \ee
   The general solution to~\eqref{gg-chi-1} is
\be \chi_{0,1} = \frac{4\xi \Lambda a}{5} (x + r_H) \int_{x_2}^x d x' \left[ \frac{(x'+r_H)}{f(x')} - \frac{(x_1 + r_H)^3}{(x'+r_H)^2 f(x')} \right]  \;, \ee
with $x_1$ and $x_2$ arbitrary constants.  Evaluating the integral near the event horizon ($x = 0$) one finds that the solutions are divergent unless $x_1 = 0$.  Evaluating near
the cosmological horizon ($x = r_C - r_H$) one finds that the solutions are divergent unless $x_1 = r_C - r_H$.  Thus for $\xi \ne 0$ there is no solution that is regular at both horizons.

\subsection{BEC acoustic black holes}
\label{sec-BEC}

This case has similarities and differences with each of the previous cases.  As shown in~\cite{abfp}, for the specific model we consider, in which $c=c(x)$ and $n,\ v$ are constant, the metric coefficients in~(\ref{geme}) are given by
$f=\frac{c^2-v^2}{c}$ and $\ C_{\perp} = \frac{1}{\sqrt{c}}$ where as stated in Sec.~\ref{sec-volterra}  we are ignoring the irrelevant constants $m$ and $n$. The condensate moves with a constant velocity $v$ in the negative $x$ direction and
it is assumed that $c > |v|$ for $x > 0$ and $c < |v|$ for $x < 0$ so there is just one horizon at $x = 0$ where $c(0) = |v|$.
As for Schwarzschild spacetime, the acoustic metric is asymptotically flat since $c$ approaches the constant $c_2>|v|$ in the limit $x \to \infty$.
The crucial difference, however, is in the behavior of the conformal factor for the transverse part of the line element $C_{\perp}$. In Schwarzschild spacetime $C_{\perp}=x+r_H$ is regular at the horizon but diverges at infinity; here $C_{\perp} = \frac{1}{\sqrt{c}}$ is regular both at the horizon and at infinity.  Thus, as in Schwarzschild-de Sitter spacetime
the graybody factor will approach a nonzero constant in the zero frequency limit.
The effective potential is 
\be V_{\rm eff}=-\frac{c}{2}(1-\frac{v^2}{c^2})^2c^{''}+\frac{1}{4}(1-\frac{v^2}{c^2})(1-\frac{5v^2}{c^2})c'^{2}\;. \label{veff-bec} \ee
It is plotted in Fig. 5 for the profile considered in~\cite{abfp}:
\be c = \sqrt{c_1^2+ \frac{1}{2}(c_2^2-c_1^2)\left[1 + \frac{2}{\pi} \tan^{-1}\left( \frac{x+d}{\sigma_v} \right) \right]} \label{c-profile} \ee
with
\be d = \sigma_v \tan\left[\frac{\pi}{c_2^2-c_1^2} \left(v_0^2-\frac{1}{2} (c_1^2+c_2^2)\right)\right] \;. \label{d-profile} \ee
There is a well with $V_{\rm eff} < 0$ located close to the horizon followed by a barrier. The well is a generic feature, since near the horizon
\be \label{hno} V_{\rm eff}\sim - (1-\frac{v^2}{c^2})c'^{\, 2}<0\ .\ee
Thus as expected from the discussion in~\ref{sec-low-freq}, $V_{\rm eff}$ changes sign in the region outside of the horizon.

\begin{figure}
\vskip -0.2in \hskip -0.4in
\begin{center}
\includegraphics[scale=0.3,angle=90,width=3.4in,clip]{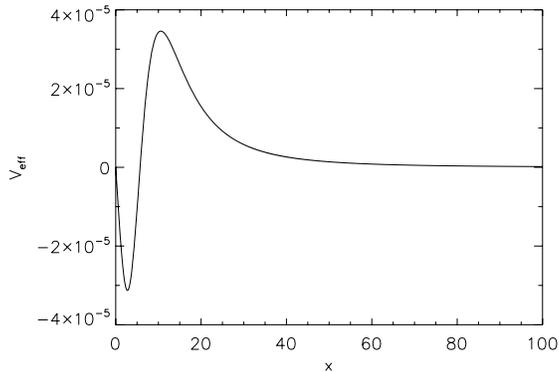}
\end{center}
\vskip -.2in \caption{Effective potential for a 1D BEC acoustic black hole with the sound speed profile~\eqref{c-profile} which was considered in \cite{abfp}, with $ \sigma_v = 8$, $|v|=\frac{3}{4}$, $c_1 = \frac{1}{2}$, and $c_2=1$.  }
\label{fig-Veff-BEC}
\end{figure}

We next proceed to determine the low-frequency scattering coefficients and gray-body
factor in this background. The general solution to the zero frequency mode equation~\eqref{chi0-general} is
\begin{equation}\label{gensol}
\chi_{0}=  \frac{a}{\sqrt{c}} + \frac{b}{\sqrt{c}} \int^x d \bar{x} \frac{c^2(\bar{x})}{c^2(\bar{x})-v^2}\ .
\end{equation}

The $\chi_0^{(1)}$ solution is constructed by noting that as $x^* \to \infty$, the second term on the right in~\eqref{gensol} has the leading order behavior $b \, \sqrt{c_2} \, x^*$.  Thus
for the condition~\eqref{chi0-1} to be satisfied $a=\sqrt{c_2}$ and $b=0$ which gives
\begin{equation}\label{chi-1-BEC}
\chi_{0}^{(1)}=\sqrt{\frac{c_2}{c(x)}}\ .
\end{equation}
Unlike (\ref{phic}), but similar to (\ref{ua}),  this solution is finite everywhere and approaches a constant, $\sqrt{c_2/|v|}$, at the horizon.
The condition~\eqref{chi0-2} for  $\chi_{0}^{(2)}$ is satisfied if $b=1/\sqrt{c_2}$ in which case
\begin{equation}\label{chis}
\chi_{0}^{(2)}= \frac{a}{\sqrt{c(x)}}+\frac{1}{\sqrt{c_2c(x)}}\int dx \frac{c^2}{c^2-v^2}\ .
\end{equation}
This solution diverges as $\sqrt{\frac{|v|}{c_2}} \,x^*$ at the horizon.
Substituting these solutions into Eq.~(\ref{ABCDsmallw3}) we obtain
\bes \begin{eqnarray}\label{cinque}
&& \mathscr{A}=  \sqrt{\frac{c_2}{|v|}} \ ,\\ \label{sei}
 && \mathscr{B}= 0\ ,\\ \label{sette}
&& \mathscr{C}= \frac{a}{\sqrt{|v|}}\ ,\\ \label{otto}
&& \mathscr{D}= \sqrt{\frac{|v|}{c_2}}\ .
\end{eqnarray} \label{ABDC-bose} \ees
Since $\mathscr{B}=0$ and $\mathscr{A} \mathscr{D} =1$, the Wronskian condition~\eqref{wron-script-ABCD} is satisfied.
Substituting~\eqref{ABDC-bose} into~\eqref{bz} and~\eqref{grbz} gives
\be \label{gaga} R_I=\frac{|v|-c_2}{|v|+c_2}+O(\omega )\ , \ R_H=\frac{c_2-|v|}{|v|+c_2}+O(\omega )\ , T_I=T_H=\frac{2\sqrt{|v|c_2}}{c_2+|v|}+O(\omega)\;.
\ee
and
\be\label{vvv}
\Gamma = \frac{4|v|c_2}{(c_2+|v|)^2}\ ,
\ee
in agreement with the result in \cite{abfp2}.
Thus $\Gamma$ approaches a constant nonzero value as $\omega\to 0$.

 Examination of the near-horizon dip of the effective potential  in Fig.\ 5 shows that it is possible that `bound states' could exist.  Such states would come from
 spatially normalized solutions characterized by a purely imaginary frequency $\omega=i\Lambda$ with $\Lambda>0$, and they would correspond to classical instabilities ($\chi\sim e^{\Lambda t}$) which grow exponentially with time.  We have found no solutions of this form for the effective potential~\eqref{veff-bec} with the profile~\eqref{c-profile}.

 To complete this subsection , we point out that for more general models in which $c,\ v$ and $n$ vary with $x$, we get the metric (\ref{geme}) in terms of the spatial coordinate $\bar x$ defined by  $d \bar{x} =  \frac{n(x)}{m} dx$, and 
$f=\frac{n}{m}\frac{(c^2-v^2)}{c}, \ C_{\perp} = \sqrt{\frac{n}{mc}}$. \footnote{  By making the substitution $\frac{n}{m}\to \rho$, where 
$\rho$ is the fluid density (see for instance \cite{blv}), this applies for other systems as well, and not just for BECs.} For 1D acoustic black holes $c,\ v $ and $n$ 
(and thus $f$ and $C_{\perp}$) always approach finite constants both at infinity and at the horizon. 
From the general solution to the $\omega=0$ mode equation (\ref{chi0-general}) we see that, since $C_\perp$ is asymptotically finite, the solution $\chi_0^{(1)}$ has  $b=0$. This implies, from (\ref{dchi0dxstar}), that $B=0$. Thus even for this general case, the gray-body factor approaches a nonzero constant  in the zero frequency limit.

\section{Infrared behaviors of correlation functions}
\label{sec-correlation}

The method of finding solutions we have developed in Secs.~\ref{sec-volterra} - \ref{sec-low-freq} also allows for a detailed analysis of the infrared behavior of correlation functions for
a BEC acoustic black hole with one spatial dimension~\cite{abfp} and for
a spherically symmetric black hole.  In the latter case we consider only the s-wave ($\ell = 0$) contributions to the correlation functions.

 The gray-body factor enters in the correlation functions through the scattering coefficients. In the region outside the event horizon, the symmetrized two-point function for the quantized massless, minimally coupled scalar field $\hat \phi$ in the Unruh state~\cite{unruh} (the state that describes Hawking's thermal emission) takes the form~\cite{abfp}
\be \frac{1}{2} \langle \{ \hat{\phi}(t,x), \hat{\phi}(t',x') \} \rangle = \frac{I+J}{2C_{\perp}(x)C_{\perp}(x')}   \;, \label{IplusJ} \ee
where
\bes \bea
 & &  I =  N^2 \int_0^\infty d \omega \frac{\left[\chi^H_\w(x) \, {\chi_\w^H}^*(x') e^{-i \w (t-{t'})} +c.c.\right]}{\sinh\left(\frac{\pi \omega}{\kappa}\right)}\ ,
 \label{Idef} \\
 & &  J = N^2 \int_0^\infty d \omega \, \left[ \chi^I_\w (x) \, {\chi_\w^I}^*(x') e^{-i \w (t-{t'})}
   +  c.c. \right] \;.  \label{Jdef}
\eea \label{IJdef} \ees
Here $\{\ ,\ \}$ stands for the anticommutator and the normalization is $N = \frac{1}{\sqrt{4 \pi}}$ for a 1D BEC and $N = \frac{1}{4 \pi}$ for a spherically symmetric
 black hole.  The mode functions  $\chi^H_\w$ and $\chi^I_\w$
are defined in Eqs.~\eqref{chiIH}.  Their asymptotic behaviors are given in Eqs.~\eqref{chi-I-inf},~\eqref{chi-I-H},~\eqref{chi-H-inf} and~\eqref{chi-H-H}.

We wish to study the infrared behaviors of the integrands of $I$ and $J$.  To do so we first use Eqs.~\eqref{chi-I-inf} and~\eqref{chi-H-inf} to write
$\chi^I_\w$ and $\chi^H_\w$ in terms of $\chi^\infty_r$ and $\chi^\infty_\ell$.  The resulting expressions for the integrands are
\bes \bea
   I_{\rm int} &=& \frac{1}{\w \sinh(\pi \w/\kappa)} \, |T_H(\w)|^2 \left[ \chi^\infty_r (x) (\chi^\infty_r (x'))^* e^{-i \w (t - t')} \right. \nonumber \\
   & &  \left. +
   (\chi^\infty_r (x))^* \chi^\infty_r (x' ) e^{i \w (t - t')}  \right]\ ,  \label{Iint-1}  \\
   J_{\rm int} &=& = \frac{1}{ \w}  \left\{ \left[\chi^\infty_\ell (x) + R_I(\w) \chi^\infty_r(x) \right] \left[(\chi^\infty_\ell (x'))^* + R^*_I(\w) (\chi^\infty_r(x'))^* \right] e^{-i \w (t - t')} \right. \nonumber \\
     & &  \left.   \;\; +  \left[(\chi^\infty_\ell (x))^* + R^*_I(\w) (\chi^\infty_r(x))^* \right] \left[\chi^\infty_\ell (x') + R_I(\w) \chi^\infty_r(x') \right] e^{i \w (t - t')} \right\}  \ . \label{Jint-1}
\eea \label{I-J-int-1} \ees
The operations of integration over the frequency and taking the limits $x^* \to \pm \infty$ do not commute.  Thus the integrals must be computed at finite values of $x^*$ and only then
can the limits $x^* \to \pm \infty$ be taken.  For fixed $x^*$ and $x'^{*}$ in the limit $\w \to 0$ one has $\w |x^*| \ll 1$ and $\w |x'^{*}| \ll 1$.
 Thus we can use Eqs.~\eqref{chirell-chi-1-2} to write the integrands in terms of $\chi_0^{(1,2)}$.  The result is
\bes \bea  I_{\rm int} &=& \frac{2\kappa}{\pi \w^2 } \, |T_H(\w)|^2 \left[ \chi^{(1)}_0 (x) \chi^{(1)}_0 (x')  + O(\w^2) \right]\ ,
     \label{Iint-2}  \\
   J_{\rm int} &=& \frac{2}{ \w}  \left\{ \chi^{(1)}_0 (x) \chi^{(1)}_0 (x')  |1 + R_I(\w)|^2  + O(\w^2)  \right\} \ .  \label{Jint-2}
\eea  \label{I-J-int-2} \ees

For Schwarzschild and Reissner-Nordstr\"om spacetimes, the low-frequency results for the scattering coefficients~\eqref{z} show that $I_{\rm int}\sim O(1)$ and $J_{\rm int}\sim O(\w)$ in the low frequency limit.  Thus all sources of infrared divergences have been eliminated in~\eqref{IJdef} including the
 factor of $\frac{1}{\omega}$ coming from the normalization of the modes and the additional factor of $\frac{1}{\omega}$ coming from the Planckian distribution  in (\ref{Idef}).

 Radically different is the situation
for both Schwarzschild-de Sitter and BEC acoustic black holes, where (\ref{ga}) and (\ref{gaga}) show that there are infrared divergences in $I_{\rm int}$ and $J_{\rm int}$, i.e. 
$I_{\rm int}\sim \frac{1}{\w^2}$ and $J_{\rm int}\sim \frac{1}{\w}$. \footnote{Note that in the Schwarzschild de Sitter and BEC cases, by combining the first of (\ref{RTB}) with (\ref{ABCD-scrip-ABCD}), we have
$R_I= const + i*O(\w)$, thus the factor of $|1 + R_I(\w)|^2$ entering in (\ref{Jint-2}) has the form $const. + O(\w^2)$.} 
One might be concerned that these infrared divergences will persist in other types of correlation functions, such as the point-split stress-energy tensor in the gravitational case
\be\label{fgh}
\langle T_{ab}(x,t;x',t')\rangle = \frac{1}{4}\Big[ (g_{\mu\alpha'}\partial_{\alpha'}\partial_\nu+g_{\nu\alpha'}\partial_{\alpha'}\partial_\mu)-g_{\mu\nu}g^{\alpha\beta'}\partial_{\alpha\beta'}\Big]
 \langle \{ \hat{\phi}(t,x), \hat{\phi}(t',x') \} \rangle \ ,
\ee
where $g_{\alpha}^{\beta'}$ are the bivectors of parallel transport (see e.\ g.~\cite{ahs})
and the density-density correlation function in the BEC case (see e.\ g.~\cite{abfp2})
\be \label{ilm}
 \langle \hat n_1(t,x)\hat n_1(t',x')\rangle = A \Big( \partial_t\partial_{t'}-v\partial_t\partial_{x'}
-v\partial_{t'}\partial_x+v^2\partial_x\partial_{x'} \Big)  \langle \{ \hat{\phi}(t,x), \hat{\phi}(t',x') \} \rangle \ ,
\ee
where $A\sim \frac{1}{c^2(x)c^2(x')}$. 
Given the structure of the mode functions, each time derivative will always bring down one factor of $\omega$, but this is not true for the spatial derivatives due to
the factors of $\frac{1}{C_{\perp}}$ in (\ref{IplusJ}) which give non vanishing $\omega$ independent contributions.
Therefore, one would expect that in (\ref{fgh}) and (\ref{ilm}) the terms containing two spatial derivatives will have the same infrared structure as (\ref{IplusJ}) which, in both Schwarzschild-de Sitter and BEC acoustic black holes, has a leading infrared divergent term $\int \frac{dw}{w^2}$.\footnote{Recall that for a 4D massless minimally coupled scalar field in Schwarzschild-de Sitter spacetime it is the $\ell=0$ mode that fixes the infrared behavior of the correlation function (the gray-body factor for $\ell\neq 0$ goes to $0$ as $\omega \to 0$, see e.\ g.~\cite{kpp}). Similarly, in the (quasi) 1D BEC case for modes with non vanishing transverse momentum the mode function $\phi$ in~\eqref{kg} (and hence also $\chi$) has a factor of the form $ e^{ik_{\perp}x_{\perp}}$.  The quantity $k_{\perp}$ acts as a mass in the 1D hydrodynamic theory and naturally regulates any infrared divergences~\cite{cfpba}.}  However, as we show next all of the IR divergent terms in~\eqref{fgh} and~\eqref{ilm} cancel leaving these correlation functions infrared finite.\footnote{This result was so unexpected that it was missed in~\cite{abfp} where it was incorrectly stated that there are infrared divergent terms in the density-density correlation function.}

For large values of $\w$ one has to be careful because the operations of differentiation of~\eqref{IplusJ} and integration over $\w$ do not necessarily commute~\cite{abfp}.  However, there
is no such problem for small values of $\w$.  To see what happens to the infrared divergences when spatial derivatives of~\eqref{IplusJ} are taken we first substitute the solutions for
$\chi_0^{(1)}$ in Scharzschild-de Sitter spacetime~\eqref{ua} and a BEC analog black hole~\eqref{chi-1-BEC}, into~\eqref{I-J-int-2} and then divide by the factor of $C_\perp(x) C_\perp(x')$ in~\eqref{IplusJ}.  Noting that $C_\perp(x)  = x + r_H$ in Schwarzschild-de Sitter and $C_\perp(x) = 1/\sqrt{c(x)}$ for the acoustic metric, we see that in both cases the factors of $C_\perp$ are canceled leaving a generic infrared structure of the two-point function of the form
\be\label{cuna}  \langle \{ \hat \phi(t,x), \hat \phi(t',x')\} \rangle = C_1\  \int \frac{d\omega}{\omega^2}\ + C_2\ \int \frac{d\omega}{\omega}\   +\ {\rm IR\ finite\ terms}\ \;, \ee
with $C_1$ and $C_2$ constants.
It is clear that acting on it with the differential operators in (\ref{fgh}) and (\ref{ilm}) will remove these IR divergences leaving an infrared finite result (in the BEC case this has been checked numerically~\cite{unpublished}).

It is also possible using~\eqref{I-J-int-1} to see that the infrared divergences in~\eqref{IplusJ} do not contribute to divergences in~\eqref{fgh} and~\eqref{ilm} near the horizon by examining the behaviors of the modes
$\chi^\infty_r$ and $\chi^\infty_\ell$ near the horizon for fixed $\w$ so that one cannot make the assumption that $\w |x^*| \ll 1$.
Because of the definitions~\eqref{chi-r-l-c-s} it is sufficient to consider $\chi^c_\w$ and $\chi^s_\w$.
  From these definitions and Eqs.~\eqref{I-J-int-1} and~\eqref{IplusJ} it is easy to see that for every term in the
integrands of~\eqref{IplusJ} there is always a factor of either $\chi^c_\w(x)/C_\perp(x)$ or $\chi^s_\w(x)/C_\perp(x)$ and a second factor of one
of these quantities evaluated at $x'$.  Further in Eqs.~\eqref{fgh} and~\eqref{ilm} there is at most one derivative with respect to $x$ and one derivative
with respect to $x'$ in each term.  Thus in the limit that we approach the horizon it is sufficient to consider an expansion of $\chi^c_\w(x)/C_\perp(x)$ or $\chi^s_\w(x)/C_\perp(x)$
to first order in $x$.

For the metrics we consider, near the horizon $C_\perp$has the general form~\eqref{Cperp-exp}.
   In Appendix B it is shown that near the horizon
\bes   \bea \label{uau}&& \chi^c_\omega = \cos\omega x^* \left[ A_\w + \alpha_1 x \frac{\kappa A_\w - \omega B_\w}{4\kappa(\kappa^2+\omega^2)} + O(x^2) \right] \nonumber \\
     & & \qquad  + \sin\omega x^* \left[ B_\w + \alpha_1 x
\frac{\kappa B_\w + \omega A_\w}{4\kappa(\kappa^2+\omega^2)} + O(x^2) \right]   \ \ \ \  \\
\label{aua} && \chi^s_\omega = \cos\omega x^* \left[ C_\w + \alpha_1 x \frac{\kappa C_\w - \omega D_\w}{4\kappa(\kappa^2+\omega^2)} + O(x^2) \right] \nonumber \\
 & &  \qquad  + \sin\omega x^* \left[ D_\w + \alpha_1 x
\frac{\kappa D_\w + \omega C_\w}{4\kappa(\kappa^2+\omega^2)} + O(x^2)\right] \; ,
\eea \label{chi-c-s-3} \ees
with
\be \alpha_1 = 4 \kappa^2 \frac{C'_\perp(0)}{C_\perp(0)}  \;. \label{alpha-1} \ee
Note that the expression for $\chi^s_\w$ can be obtained from that for $\chi^c_\w$ via the substitutions $A_\w \to C_\w$ and $B_\w \to D_\w$.  So we focus
on $\chi^c_\w/C_\perp$.  Combining~\eqref{Cperp-exp} and~\eqref{uau} and using~\eqref{alpha-1} one finds for all values of $\w$
\bea \frac{\chi^c_\w(x)}{C_\perp(x)}  &=& \frac{1}{C_\perp (0)}\{ (A_\w \cos \w x^*  + B_\w \sin \w x^*) \left[ 1 - \frac{C'_\perp (0)}{C_\perp (0)} \frac{\w^2}{\kappa^2 + \w^2} x + O(x^2) \right] \nonumber \\
         &  &  \qquad + (A_\w \sin \w x^*  - B_\w \cos \w x^*) \left[ \frac{C'_\perp (0)}{C_\perp (0)} \frac{ \kappa \w}{\kappa^2 + \w^2} x  + O(x^2) \right] \} \label{chic-expansion} \;. \eea
Thus terms that after differentiation with respect to $x$ survive in the limit $x \to 0$ are all multiplied by at least one factor of $\w$.  So there are no infrared divergences
in~\eqref{fgh} and~\eqref{ilm} in the near horizon limit.

For completeness we display expressions for $\chi_\omega^H$ and  $\chi_\omega^I$ in the near horizon region in the
Schwarzschild-de Sitter and BEC cases for small values of $\w$ but with no assumptions about the value of $\w |x^*|$.  For Schwarzschild-de Sitter spacetime $C_\perp$ = $x + r_H$.  Using Eqs.~\eqref{ABCD-scrip-ABCD},
\eqref{ABCD-SdS},~\eqref{ga},~\eqref{chi-r-l-c-s},~\eqref{chi-I-H} and~\eqref{chi-H-H}, we find\footnote{Note that for Schwarzschild-de Sitter spacetime and the BEC acoustic black hole it is possible to show there are
no terms of $O(x^2)$ and higher in these expressions which are not also multiplied by some positive power of $\w$.  This can be done by first fixing the value of $x$ to be small but nonzero and taking the limit $\w \to 0$ in which case one finds that $\chi^\infty_r \,, \,\chi^\infty_\ell \to \lim_{\w \to 0} \chi^c_\w = \chi^{(1)}_0$.  Next one takes the limit $\w \to 0$
 in~\eqref{chic-expansion} and compares with the exact expressions for $\chi^{(1)}_0(x)/C_\perp(x)$ using Eq.~\eqref{ua} for Schwarzschild de Sitter and~\eqref{chi-1-BEC} for the BEC case.}
\bea
&& \frac{\chi_\omega^H (x)}{x+r_H} = \frac{1}{r_H} \left\{ e^{i\omega x^*} \left[1 +O(\w x) \right] + R_H(\w = 0) e^{-i\omega  x^*} \left[1+ O(\w) + O(\w x) \right] \right\}
\ , \nonumber \\
&& \frac{\chi_\omega^I}{x+r_H} = T_I(\omega=0) \frac{e^{-i\omega  x^*}}{r_H} \left[1 + O(\w) + O(\w x) \right]  
 \;.
\label{z1}
\eea
For the BEC acoustic metric $C_\perp = 1/\sqrt{c(x)}$.  Using Eqs.~\eqref{ABCD-scrip-ABCD}, \eqref{ABDC-bose},~\eqref{gaga},~\eqref{chi-r-l-c-s},~\eqref{chi-I-H} and~\eqref{chi-H-H}, we find
\bea
&&
\sqrt{c} \chi_\omega^H = \sqrt{|v|} \left\{ e^{i\omega x^*} \left[1 +O(\w x) \right] + \sqrt{|v|} R_H(\w = 0) e^{-i\omega  x^*} \left[ 1+O(\w) +O(\w x) \right] \right\}   \;,
\nonumber \\
&&
\sqrt{c} \chi_\omega^I = \sqrt{|v|}  T_I(\omega=0) e^{-i\omega  x^*} \left[1 + O(\w) + O(\w x) \right] 
 \;.
 \label{zz1}\eea

\section{Conclusions}
\label{sec-concl}

In this paper we have developed a rigorous method based on Volterra integral equations of the second kind~(\ref{volterra}) to determine the low-frequency behaviors of the scattering coefficients and gray-body factor for a massless minimally coupled scalar field satisfying the Klein-Gordon equation (\ref{kg}) in the static metric (\ref{geme}).  These are given in (\ref{RTB}), (\ref{RTH}) and (\ref{grb}), in terms of coefficients~\eqref{ABCD-scrip-ABCD} and~\eqref{ABCDsmallw3} that can be expressed as horizon boundary values of the real solutions
to the zero frequency mode equation $\chi_0^{(1,2)}$ with the asymptotic behaviors~\eqref{chi0-1-2}.
  These results are valid for the modes of the scalar field which are either in the zero angular momentum or s-wave sector for a spherically symmetric black hole, or in the longitudinal sector of a BEC acoustic black hole with effectively one spatial dimension. 
 
 From the general analysis of Sec.~\ref{sec-low-freq} we see that there are two qualitatively different behaviors for the gray-body factor which depend on
 the value of the coefficient $\mathscr{B}$ in (\ref{ABCDsmallw3}).
 If $\mathscr{B} \neq 0$  the gray-body factor vanishes as $\omega^2$.  As shown in Sec.~\ref{sec-sch-rn} this occurs for Schwarzschild and Reissner-Nordstr\"om black holes.  If $\mathscr{B}=0$  the gray-body factor approaches a nonzero constant as $\w \to 0$.  As shown in Secs.~\ref{sec-schds} and~\ref{sec-BEC} this occurs  for Schwarzschild-de Sitter black holes and a 1D BEC acoustic black hole.  It was shown in Secs.~\ref{sec-low-freq} and~\ref{sec-schds} that a necessary, but not sufficient condition to have $\mathscr{B} \ne 0$ is for the effective potential, $V_{\rm eff}$ to change sign.  It was also shown in Sec.~\ref{sec-low-freq}
 that a necessary condition to have $\mathscr{B} \ne 0$ is that the zero frequency solution $\chi^{(1)}_0$ be bounded at the event horizon.  By definition~\eqref{chi0-1}, $\chi^{(1)}_0$ is always bounded either at the cosmological horizon or infinity for the cases we consider.

  The results for the scattering coefficients and gray-body factor were used to study infrared divergences.  The constant gray-body factor at small $\omega$ for both Schwarzschild-de Sitter and 1D BEC acoustic black holes implies that Hawking radiation is dominated by an infinite number of low energy particles as can be seen from Eq.~\eqref{kjk}.  As shown
  in Sec.~\ref{sec-correlation} it also implies that the two-point function (\ref{IplusJ})
 is infrared divergent with the general form of the divergences given in~(\ref{cuna}).  No such divergences exist in Schwarzschild and Reissner-Nordstr\"om spacetimes.  Despite this fact, we proved that, both away from the horizon and close to it, the point-split stress-energy tensor (\ref{fgh}) in Schwarzschild-de Sitter spacetime and the density-density correlation function (\ref{ilm}) for the 1D BEC acoustic black hole are always infrared finite.

\begin{acknowledgments}  We thank Renaud Parentani for useful discussions.  
This work was supported in part by the National Science Foundation under Grant Nos. PHY-0856050 and PHY-1308325.
Some of the plots were generated using the WFU DEAC cluster;
we thank WFU's Provost Office and Information Systems Department for their generous support. 
\end{acknowledgments}

\vspace{1cm}

\appendix

\section{Some results used for the computation of $A_\w$} 

\vspace{0.3cm}

In this appendix we give several proofs relating to the IR behaviors of the
coefficients in~\eqref{abcddef}.  Throughout we focus on $A_\w$.  The generalization of the proofs
to the other coefficients is straight-forward.

We begin by writing the integral in~\eqref{abcddef} for $A_\w$ in the form
\bea A_\w &=& 1 + I_1 + I_2 + I_3 \nonumber \\
    I_1 & = & \frac{1}{\w} \int_{-\infty}^{-\Lambda} dy^* \sin(\w y^*) V_{\rm eff}(y^*)  \chi^c_\w(y^*) \nonumber \\
    I_2 &=& \frac{1}{\w} \int_{-\Lambda}^{\Lambda} dy^* \sin(\w y^*) V_{\rm eff}(y^*)  \chi^c_\w(y^*) \nonumber \\
  I_3 &= &   \frac{1}{\w} \int_{\Lambda}^{\infty} dy^* \sin(\w y^*) V_{\rm eff}(y^*)  \chi^c_\w(y^*) \;.  \label{A-Lambda}
\eea
The first proof involves showing that the first integral vanishes in the limit $\Lambda \rightarrow \infty$ for all $\w$ including
$\w = 0$ if we take the limits $\w \to 0$ and $\Lambda \to \infty$ in such a way that $\w \Lambda \ll 1$.  Note that this condition on the limits allows an expansion of the oscillatory functions in the integrand of the middle integral in powers of $\w y^*$ and also allows one to expand $V_{\rm eff}$ in the limit of large negative and large positive $y^*$ in the first and third integrals respectively.

Since the first integral in~\eqref{A-Lambda} covers the region near the horizon it is useful to rewrite the general form of the Volterra equations~\eqref{volterra} for $\chi^c_\w$ and $\chi^s_\w$ so that the integration range is $-\infty < y^* \le x^*$.  This can be done by substituting~\eqref{abcd} into~\eqref{volterra} and using~\eqref{abcddef} and~\eqref{ABCDdef}.  The result is
\bes
\bea
 \chi^c_\w &=&  A_\w \cos \w x^*  \, + B_\w \sin \w x^*  \, + \frac{1}{\w} \int_{- \infty}^{x^*} dy^* \sin[\w (x^* - y^*)]  V_{\rm eff}(y^*) \, \chi^c_\w (y^*)\ ,  \label{chic-exp} \\
 \chi^s_\w &=&  C_w \cos \w x^*  \, + D_\w \sin \w x^*  \, + \frac{1}{\w} \int_{- \infty}^{x^*} dy^* \sin[\w (x^* - y^*)]  V_{\rm eff}(y^*) \, \chi^s_\w (y^*) \;. \label{chis-exp}
\eea
\label{chic-chis}
\ees

We bound the first integral in~\eqref{A-Lambda} by bounding the integrand so that
\be |I_1| = \bigg| \frac{1}{\w} \int_{-\infty}^{-\Lambda} dy^* \sin(\w y^*) V_{\rm eff}(y^*)  \chi^c_\w(y^*) \bigg| \le  \frac{1}{\w} \int_{-\infty}^{-\Lambda} dy^* |V_{\rm eff}(y^*)|  \, |\chi^c_\w(y^*)|  \;.  \label{I1-bound} \ee
Using~\eqref{chic-chis} and iterating we find
\bea
 \chi^c_\w(x^*) &=&  A_\w \cos \w x^*  \, + B_\w \sin \w x^*  \, + \sum_{n=1}^\infty \frac{1}{\w^n}  \int_{-\infty}^{x^*} dy^*_1   \, ...  \int_{-\infty}^{y^*_{n-1}} dy^*_n \sin[\w (x^*-y^*_1)] \nonumber \\
 && \qquad \times V_{\rm eff}(y^*_1) ... \sin[\w (y^*_{n-1} - y^*_n)] V_{\rm eff}(y_n^*) \left( A_\w \cos \w y^*_n  \, + B_\w \sin \w y^*_n \right) \;.
\eea
Thus
\bea
|\chi^c_\w(x^*)| &\le&  |A_\w|  \, + |B_\w|  \, + \sum_{n=1}^\infty \frac{1}{\w^n}  \int_{-\infty}^{x^*} dy^*_1   \, ...  \int_{-\infty}^{y^*_{n-1}} dy^*_n
 |V_{\rm eff}(y^*_1)| ... |V_{\rm eff}(y_n^*)| \left( |A_\w|  \, + |B_\w| \right) \;,  \nonumber \\
  &=& (|A_\w|  \, + |B_\w|) \left( 1 + \, \sum_{n=1}^\infty \frac{1}{n! \w^n}  \int_{-\infty}^{x^*} dy^*_1   \, ...  \int_{-\infty}^{x^*} dy^*_n
 |V_{\rm eff}(y^*_1)| ... |V_{\rm eff}(y_n^*)|  \right) \;, \nonumber \\
 &=& (|A_\w|  \, + |B_\w|) \exp \left( \frac{1}{\w} \, \int_{-\infty}^{x^*} dy_1^* |V_{\rm eff}(y_1^*)| \right)  \;.  \label{chic-bound}
\eea
We want to use this bound in~\eqref{I1-bound}.  Since the integrand in~\eqref{I1-bound} covers the range $-\Lambda \ge y^* > -\infty$ it is possible to make the further bound
\be |\chi^c_\w(y^*)| \le (|A_\w|  \, + |B_\w|) \exp \left( \frac{1}{\w} \, \int_{-\infty}^{-\Lambda} dy_1^* |V_{\rm eff}(y_1^*)| \right) \;. \ee
Then~\eqref{I1-bound} becomes
\be |I_1| <  (|A_\w|  \, + |B_\w|) \exp \left( \frac{1}{\w} \, \int_{-\infty}^{-\Lambda} dy_1^* |V_{\rm eff}(y_1^*)| \right) \;  \frac{1}{\w} \int_{-\infty}^{-\Lambda} dy^* |V_{\rm eff}(y^*)| \;. \label{I1-bound2} \ee

This shows that $I_1$ vanishes for all $\w >0$ in the limit $\Lambda \rightarrow \infty$.  For the limit $\w \to 0$ with $\w \Lambda \ll 1$
we can take the limits in such a way that $\w \Lambda^{1+\epsilon} = 1 $, for some $\epsilon > 0$.  Then
a sufficient condition that $I_1$ should vanish for $\Lambda \rightarrow \infty$ is that
\be \Lambda^{1+\epsilon} \int_{-\infty}^{-\Lambda} dy^* |V_{\rm eff}(y^*)| \rightarrow 0  \;. \label{condition-1}\  \ee

To go further we need to find a more explicit expression for $V_{\rm eff}$ near the horizon.  This can
be done by writing it in terms of a power series in $x$.
Expansions for $f$ and $C_\perp$ in powers of $x$ have been given in Secs.~\ref{sec-volterra} and~\ref{sec-correlation} respectively.
Using~\eqref{veff} along with~\eqref{f-exp},~\eqref{f-exp-2}, and~\eqref{Cperp-exp} one finds
\be V_{\rm eff} = \sum_{i = 1}^\infty \alpha_i x^i \;, \ee
with
\be \alpha_1 = 4 \kappa^2 \frac{C'_\perp(0)}{C_\perp(0)}  \;. \label{alpha-1-2} \ee
Using the definition~\eqref{xstar-def} along with the expansion~\eqref{f-exp} for $f$ one can show
by iteration that near the horizon
\be x(x^*) = \sum_{k=1}^\infty \beta_k e^{2 k \kappa x^*} \;. \label{x-xstar} \ee
Thus it is also possible to write
\be V_{\rm eff} = \sum_{j=1}^\infty \gamma_j e^{2 j \kappa x^*} \;,  \label{veff-hor-xstar}\ee
with
\be \gamma_1 = \alpha_1 \beta_1 \;. \label{gamma-1}  \ee
If we bound each term in this last expansion and substitute into~\eqref{condition-1} then it is easy to show that this condition is satisfied for any $\epsilon >0$
so long as the resulting sum converges.\footnote{Even if the sum is an asymptotic series, each term in this series vanishes in the limit $\Lambda \to \infty$.}

We next bound the third integral in~\eqref{A-Lambda} by bounding its integrand so that
\be |I_3| = \bigg| \frac{1}{\w} \int_{\Lambda}^{\infty} dy^* \sin(\w y^*) V_{\rm eff}(y^*)  \chi^c_\w(y^*) \bigg| \le  \frac{1}{\w} \int_{\Lambda}^{\infty} dy^* |V_{\rm eff}(y^*)|  \, |\chi^c_\w(y^*)|  \;.  \label{I3-bound} \ee
Using Eq.~\eqref{volterra} and iterating we find that
\bea  &&\chi^c_\w(x^*) = \cos(\w x^*)  +  \sum_{n=1}^\infty  \frac{(-1)^n}{\w^n} \int_{x^*}^\infty dy^*_1 ... \int_{y_{n-1}^*}^\infty dy^*_n \sin[\w (x^*-y^*_1)] \nonumber \\
& & \qquad \times \, V_{\rm eff}(y_1^*) \, ... \, \sin[\w(y^*_{n-1}-y^*_{n})] \, V_{\rm eff}(y_n) \, \cos(\w y^*_n) \;. \label{chic-total} \eea
Thus
\be |\chi^c_\w (x^*)| \le \exp \left( \frac{1}{\w} \int_{x^*}^\infty d y_1^* |V_{\rm eff}(y_1^*)|  \right) \;. \label{chic-upper-bound} \ee
We want to use this bound in~\eqref{I3-bound}.  Since the integrand in~\eqref{I3-bound} covers the range $\Lambda \le y^* < \infty$ it is possible to make the further bound
\be |\chi^c_\w (x^*)| \le \exp \left( \frac{1}{\w} \int_{\Lambda}^\infty d y_1^* |V_{\rm eff}(y_1^*)|  \right) \;. \label{chic-upper-bound-2} \ee
Substituting into~\eqref{I3-bound} one finds
\be |I_3| \le  \exp \left( \frac{1}{\w} \int_{\Lambda}^\infty d y_1^* |V_{\rm eff}(y_1^*)|  \right) \; \frac{1}{\w} \int_\Lambda^\infty d y |V_{\rm eff}(y^*)|   \;. \label{I3-bound2} \ee
This clearly vanishes for all $\w >0$ in the limit $\Lambda \rightarrow \infty$.
For the limit $\w \to 0$ with $\w \Lambda \ll 1$, as above
we can take the limits in such a way that $\w \Lambda^{1+\epsilon} = 1 $, for some $\epsilon > 0$.  Then
a sufficient condition that $I_3$ should vanish for $\Lambda \rightarrow \infty$ is that
\be \Lambda^{1+\epsilon} \int_{\Lambda}^{\infty} dy^* |V_{\rm eff}(y^*)| \rightarrow 0 \;. \label{condition-2} \ee
In this case however there are restrictions on the value of $\epsilon$ which relate to the asymptotic behaviors of $V_{\rm eff}$.
For Schwarzschild and Reissner-Nordstr\"om spacetimes and for a BEC acoustic black hole with sound speed profile~\eqref{c-profile} it
is not difficult to show that for large $x$, $V_{\rm eff} \sim x^{-3}$.  In this case any value of $\epsilon$ in the range $0 < \epsilon < 1$
will work.  For Schwarzschild-de Sitter spacetime it is not hard to show that
$V_{\rm eff} \sim e^{-2 \kappa_C x^*}$ with $\kappa_C$ the surface gravity of the cosmological horizon.  In this case the only restriction on $\epsilon$ is
that it be positive.

Next we show how to manipulate $I_2$ to obtain~\eqref{ABCDdefstep1}.
Using the mode equation~\eqref{schreq} one finds that
\be I_2 =\frac{1}{\omega} \int_{-\Lambda}^\Lambda dy^* \sin(\w y^*) \left( \frac{d^2 \chi^c_\w(y^*)}{(d y^*)^2}  + \w^2 \chi^2_\w (y^*) \right)   \;. \ee
Integrating the first term by parts gives
\be I_2 = \left[ \frac{\sin(\w y^*)}{\w} \frac{d \chi^c_\w(y^*)}{d y^*} \right]_{-\Lambda}^\Lambda  + \frac{1}{\w} \int_{-\Lambda}^\Lambda dy^* \left[ - \w \cos(\w y^*) \frac{d \chi^c_\w(y^*)}{d y^*} + \w^2 \sin(\w y^*) \chi^c_\w(y^*) \right]  \;. \ee
Integrating a second time by parts gives
\be I_2 = \left[ \frac{\sin(\w y^*)}{\w} \frac{d \chi^c_\w(y^*)}{d y^*} - \cos(\w y^*) \chi^c_\w(y^*) \right]_{-\Lambda}^\Lambda \;. \ee
One can then take the limit $\Lambda \rightarrow \infty$ if either $\w >0 $ or if the limit $\w \rightarrow 0$ is taken such that $\w \Lambda \ll 1$.

 Finally we argue that $\mathscr{A} = A + O(\w^2)$.  First note that from Eq.~\eqref{veff} it is clear that so long as $C_\perp$ is analytic in $x$ and nonvanishing then $V_{\rm eff}$ is as well.  Further, $x^*(x)$ is also analytic except at $x = 0$.  Thus $\chi^c_\w$ is analytic both in $x^*$ and $\w$ since, as can be seen from~\eqref{chic-total}, the integrand for the integral over $y_i^*$ is analytic in $y^*_i$, $y^*_{i-1}$, and $\w$.  This is correct even in the limit $\w \rightarrow 0$.
Next note from~\eqref{chic-total} that $\chi^c_\w$ is an even function of $\w$.
Then from~\eqref{abcddef} it is clear that $A_\w$ is also an analytic function of $\w$ and an even function of $\w$.  Thus $A_\w = \mathscr{A} + O(\w^2)$.

\section{Near-horizon computations of $\chi^c_\w$ and $\chi^s_\w$}

In this appendix we use the Volterra equation to derive the exact near-horizon behaviors of the solutions $\chi^c_\w$ and $\chi^s_\w$.
We begin by substituting the expansion~\eqref{veff-hor-xstar} into~\eqref{chic-exp} with the result that
\bea
 \chi^c_\w(x^*) &=&  A_\w \cos \w x^*  \, + B_\w \sin \w x^*  \, + \sum_{n=1}^\infty \left\{ \sum_{j_1=1}^\infty ... \sum_{j_n = 1}^\infty \frac{\gamma_{j_1} ... \gamma_{j_n}}{\w^n} \int_{-\infty}^{x^*} dy^*_1   \, ...  \int_{-\infty}^{y^*_{n-1}} dy^*_n \right. \nonumber \\
 & & \left. \times  \sin[\w(x^*-y^*_1)] ... \sin[\w(y^*_{n-1} - y^*_n)] e^{2 j_1 \kappa y^*_1} ... e^{2 j_n \kappa y^*_n} \right. \nonumber \\
   & & \left. \times \left(A_\w \cos \w y^*_n  \, + B_\w  \sin \w y^*_n \right) \right\} \;.
 \eea
Examination of~\eqref{chis-exp} shows that the corresponding expansion for $\chi^s_\w$ is obtained with the substitutions $A_\w \rightarrow C_\w$ and $B_\w \rightarrow D_\w$.

At this point the integrals can all be computed starting with the integral over $y^*_n$ and working in order to the integral over $y^*_1$.  The first integral is
\bea & & \int_{-\infty}^{y^*_{n-1}} dy^*_n \sin[\w(y^*_{n-1} - y^*_n)] e^{2 j_n \kappa y^*_n} \left( A_\w \cos \w y^*_n  \, + B_\w \sin \w y^*_n \right)
  \nonumber \\
 & &  \qquad \qquad = \w \exp(2 j_n \kappa y^*_{n-1}) \left( A_\w^{(1)} \cos \w y^*_{n-1}  \, + B_\w^{(1)} \sin \w y^*_{n-1} \right) \;,
 \eea
 with
 \bes
 \bea A_\w^{(1)} &=& \frac{ 2 j_n \kappa A_\w - 2 \w B_\w }{(2 j_n \kappa)^3 + 4 (2 j_n \kappa) \w^2} \ , \\
  B_\w^{(1)} &=& \frac{ 2 j_n \kappa B_\w + 2 \w A_\w }{(2 j_n \kappa)^3 + 4 (2 j_n \kappa) \w^2}\ .
\eea
\label{A-1-B-1}
\ees
The next integral is then
\bea & & \int_{-\infty}^{y^*_{n-2}} dy^*_{n-1} \sin[\w(y^*_{n-2} - y^*_{n-1})] e^{2 (j_{n-1}+j_n) \kappa y^*_n} \left( A^{(1)}_\w \cos \w y^*_{n-1}  \, + B^{(1)}_\w \sin \w y^*_{n-1} \right)
 = \nonumber \\
 & & \qquad  \qquad \w \exp(2 (j_{n-1}+j_n) \kappa y^*_{n-2})   \left( A^{(2)}_\w \cos \w y^*_{n-2}  \, + B^{(2)}_\w \sin \w y^*_{n-2} \right) \;,
\eea
with $A^{(2)}_\w$ and $B^{(2)}_\w$ obtained by making the substitutions $A^{(1)}_\w \rightarrow A^{(2)}_\w$ and $B^{(1)}_\w \rightarrow B^{(2)}_\w$ in~\eqref{A-1-B-1} followed by $A_\w \rightarrow A^{(1)}_\w$ and $B_\w \rightarrow B^{(1)}_\w$.
Clearly the same form holds for all of the integrals and the result is
 \bes
 \bea \chi^c_\w &=&  A_\w \cos \w x^*  \, + B_\w \sin \w x^*  \, + \sum_{n=1}^\infty \left\{ \sum_{j_1=1}^\infty ... \sum_{j_n = 1}^\infty \gamma_{j_1} ... \gamma_{j_n}
 \exp(2 \kappa (j_1 + ... + j_n) x^*) \right. \nonumber \\  & & \left. \right. \nonumber \\
 & & \left. \qquad \times \left( A^{(n)}_\w \cos \w x^*  \, + B^{(n)}_\w \sin \w x^* \right) \right\}\ ,  \\ & &  \nonumber \\
 \chi^s_\w &=&  C_\w \cos \w x^*  \, + D_\w \sin \w x^*  \, + \sum_{n=1}^\infty \left\{ \sum_{j_1=1}^\infty ... \sum_{j_n = 1}^\infty \gamma_{j_1} ... \gamma_{j_n}
 \exp(2 \kappa (j_1 + ... + j_n) x^*) \right. \nonumber \\   & & \left. \right. \nonumber \\
 & & \left. \qquad \times \left( C^{(n)}_\w \cos \w x^*  \, + D^{(n)}_\w \sin \w x^* \right) \right\} \ .
\eea   \label{chi-horizon}
\ees
If we include terms of $O(x)$ but not higher then using~\eqref{x-xstar} and~\eqref{gamma-1}
\bes
 \bea \chi^c_\w &=&  A_\w \cos \w x^*  \, + B_\w \sin \w x^*  \, + \alpha_1 x
  \left( A^{(1)}_\w \cos \w x^*  \, + B^{(1)}_\w \sin \w x^* \right) \ ,  \\ & &  \nonumber \\
 \chi^s_\w &=&  C_\w \cos \w x^*  \, + D_\w \sin \w x^*  \, +  \alpha_1 x
      \left( C^{(1)}_\w \cos \w x^*  \, + D^{(1)}_\w \sin \w x^* \right) \ .
\eea   \label{chi-horizon-1}
\ees

{}

\vspace{1cm}

\begin{thebibliography}{99}
\bibitem{Hawking:1974sw}
S.~W.~Hawking, Nature {\bf 248}, 30 (1974);
Commun.\ Math.\ Phys.\  {\bf 43}, 199 (1975)
\bibitem{hawkell}  S-W. Hawking and G.F.R. Ellis, {\it The large scale structure of space-time}, Cambridge University Press, Cambridge, 1973
\bibitem{teukpr} W.H. Press and S.A. Teukolsky, {\it Astrophys. J.} {\bf 193}, 443 (1974)
\bibitem{starob} A.A. Starobinsky and S.M. Churilov, {\it Sov. Phys. JETP} {\bf 38}, 1 (1974)
\bibitem{page}
D.N. Page, {\it Phys. Rev.} {\bf D13}, 98 (1976)
\bibitem{uno}
T.~Harmark, J.~Natario and R.~Schiappa,
  Adv.\ Theor.\ Math.\ Phys.\  {\bf 14}, 727 (2010)
  \bibitem{due}
  R.~Jorge, E.~S.~de Oliveira and J.~V.~Rocha,
  arXiv:1410.4590 [gr-qc]
\bibitem{uu}
S.R. Das, G. Gibbons and S.D. Mathur, {\it Phys. Rev. Lett.} {\bf 78}, 417 (1997)
\bibitem{uuu}
A. Higuchi, {\it Class. Quant. Grav.} {\bf 18}, L139 (2001) ; {\bf 19}, 599 (2002)
\bibitem{unoref}
  P.~R.~Brady, C.~M.~Chambers, W.~Krivan and P.~Laguna,
  Phys.\ Rev.\ D {\bf 55}, 7538 (1997)
  \bibitem{kgb}
P. Kanti, J. Grain and A. Barrau, {\it Phys. Rev.} {\bf D71}, 104002 (2005)
\bibitem{chor}
L.C.B. Crispino, A. Higuchi, E.S. Oliveira and J.V. Rocha, {\it Phys. Rev.} {\bf D87} 10, 104034 (2013)
\bibitem{kpp}
 P. Kanti, T. Pappas and N. Pappas, arXiv:14098664 [hep-th]
\bibitem{abfp2}
P. Anderson, R. Balbinot, A. Fabbri and R. Parentani, arXiv:1404.3224 [gr-qc], {\it Phys. Rev.} D{\bf 90}, 104044 (2014)

\bibitem{volterra-eq} Encyclopedia of Mathematics, http:$//$www.encyclopediaofmath.org$/$index.php$/$Volterra\_equation

\bibitem{unruh}  W.~G.~Unruh,
  Phys.\ Rev.\  {\bf D14}, 870 (1976)

 \bibitem{abfp}
P.R. Anderson, R. Balbinot, A. Fabbri and R. Parentani, {\it Phys. Rev.} {\bf D87}, 124018 (2013)
\bibitem{candelas}
P. Candelas, {\it Phys. Rev.} {\bf D21}, 2185 (1980)

\bibitem{dueref} P.~R.~Brady, C.~M.~Chambers, W.~G.~Laarakkers and E.~Poisson,
  Phys.\ Rev.\ D {\bf 60}, 064003 (1999)

\bibitem{blv}
C.~Barcelo, S.~Liberati and M.~Visser,
  Living Rev.\ Rel.\  {\bf 8}, 12 (2005)
  [Living Rev.\ Rel.\  {\bf 14}, 3 (2011)]
\bibitem{ahs}
P.R. Anderson, W. A. Hiscock and D.A. Samuel, {\it Phys. Rev.} {\bf D51}, 4337  (1995)
 \bibitem{cfpba}
 A. Coutant, A. Fabbri,  R. Parentani, R. Balbinot and P.R. Anderson, {\it Phys. Rev.} {\bf D86}, 064022 (2012)

 \bibitem{unpublished} P. R. Anderson, unpublished
 \end{thebibliography}
 \end{document}